\documentclass[extra, onecolumn]{gji}

\title[Influence of a stably stratified layer on the Earth's outer core waves]{The influence of a stably stratified layer on the hydromagnetic waves in the Earth's core and their electromagnetic torques.}
\author[F. Seuren \emph{et al.}]{
Fleur Seuren,$^1$$^2$
Santiago A. Triana,$^1$
Jérémy Rekier,$^1$ 
Véronique\and  Dehant$^1$$^3$  and
Tim Van Hoolst$^1$$^2$ 
\\
$^{1}$ Royal Observatory of Belgium, Ringlaan \emph{3}, BE-\emph{1180} Brussels, Belgium. 
\\
$^{2}$ Institute of Astronomy, KU Leuven, Celestijnenlaan \emph{200D}, BE-\emph{3001} Leuven, Belgium
\\
$^{3}$ Earth and Life Institute, Université Catholique de Louvain, Place Louis Pasteur \emph{3}, BE-\emph{1348} Louvain-la-Neuve, Belgium
}
\date{Accepted 2025 October 26. Received 2025 October 22; in original form 2024 December 20}
\pagerange{??--??}
\volume{???}
\pubyear{????}

\newcommand{\bb}[1]{\boldsymbol{\mathbf{#1}}}
\newcommand{\citepfull}[1]{(\citefullauthor{#1} \citeyear{#1})}
\newcommand{\citetfull}[1]{\citefullauthor{#1} (\citeyear{#1})}

\usepackage[addedmarkup=bf, todonotes={textsize=tiny}, draft]{changes}

\usepackage{amsfonts}
\usepackage{amssymb}
\usepackage[fleqn]{amsmath}
\DeclareMathOperator{\sgn}{sgn}

\usepackage{xcolor}
\usepackage{hyperref}
\usepackage{graphicx}

\begin{document}
\label{firstpage}

\maketitle

\begin{summary}
Evidence from seismic studies, mineral physics, thermal evolution models and geomagnetic observations is inconclusive about the presence of a stably stratified layer at the top of the Earth's fluid outer core. Such a convectively stable layer could have a strong influence on the internal fluid waves propagating underneath the core-mantle boundary (CMB) that are used to probe the outermost region of the core through the wave interaction with the geomagnetic field and the rotation of the mantle. Here, we numerically investigate the effect of a top stable layer on the outer core fluid waves by calculating the eigenmodes in a neutrally stratified sphere permeated by a magnetic field with and without a top stable layer. We use a numerical model, assuming a flow with an m-fold azimuthal symmetry, that allows for radial motions across the lower boundary of the stable layer and angular momentum exchanges across the CMB through viscous and electromagnetic coupling. On interannual time-scales, we find torsional Alfvén waves that are only marginally affected by weak to moderate stratification strength in the outer layer. At decadal time-scales similarly weak stable layers promote the appearance of waves that propagate primarily within the stable layer itself and resemble Magneto-Archimedes-Coriolis (MAC) waves, even though they interact with the adiabatic fluid core below. These waves can exert viscous and electromagnetic torques on the mantle that are several orders of magnitude larger than those in the neutrally stratified case.
\end{summary}


\section{Introduction}  \label{sec:intro}
Approximately 2900 kilometres below the Earth's surface lies the Earth's outer core, a vast ocean of hot molten metal alloys. Because of its deep location, all current information on the internal properties, composition, location and dynamics of this liquid layer have been indirectly inferred from seismology, high-pressure experiments and models, and observations of the Earth's magnetic field and rotation \citep[see e.g.][for respective reviews]{souriau2015deep, dehant2022structure, lesur2022rapid, rekier2021earth}, leaving many open questions. A particular area of uncertainty on the outer core structure that we will address here concerns the region just below the core-mantle boundary (CMB), which might be stable against convection. 

First indications that the outermost core region was different from the bulk liquid core were obtained at the end of the last century when seismological studies started to report anomalous seismic observations in that region (see e.g. \citeauthor{hales1971velocities} \citeyear{hales1971velocities}; \citefullauthor{garnero1993constraining} \citeyear{garnero1993constraining}), suggesting a local lower seismic velocity and lower density than the deep regions. \citep[][]{lay1990stably}. Attempts to interpret this low-velocity, low-density layer in terms of core composition turned out to be a complex endeavour, resulting in some conflicting views on the outer core structure. Some studies, for example by \citetfull{irving2018seismically}, explain the low velocities near the CMB by assuming that the same light and compressible alloy is present everywhere in the outer core. This facilitates a lower density at the top of the core due to the lower compression by the lower pressure there and causes a steep velocity gradient throughout the core. Other studies \citep[e.g.][]{kaneshima2018array} favour a less compressible material that cannot accommodate such a steep velocity gradient and require a distinct compositionally stratified layer at the top of the core. 

The existence of a thermally stratified layer is also supported by some theoretical \citep[e.g.][]{pozzo2012thermal} and experimental \citep[e.g.][]{zhang2021thermal} calculations of the thermal properties of liquid iron and iron alloys at Earth's outer core conditions. The high thermal conductivity obtained in these studies, combined with estimates of the CMB heat flow obtained by thermal evolution studies, indicates that the uppermost region of the core is subadiabatic and thus thermally stratified. Other studies, however, suggest that the conductivity of iron and iron alloys could be several times smaller than the highest estimates \citep[see e.g.][]{hsieh2020low, konopkova2016direct}. The heat conducted along an adiabat at the CMB could then be smaller than estimates of the CMB heat flux, and a thermally stratified layer at the top of the core is not needed.

Observations of the Earth's magnetic field offer no satisfying conclusion on the existence of stratification near the CMB either. A stable layer might be necessary to explain some very long millennial time-scale characteristics of the observed geomagnetic field, e.g. its zonal octupolar component \citep[see][]{yan2018sensitivity} or polarity reversals \citep[see e.g.][]{aubert2025core}. Yet, features that occur on shorter, decadal, time-scales, such as the intense geomagnetic flux patches that are presumably caused by up- and downwellings of the core fluid, are hard to reconcile with a top stable layer (see e.g. \citeauthor{amit2014can} \citeyear{amit2014can}, \citefullauthor{huguet2018geomagnetic} \citeyear{huguet2018geomagnetic}) unless that layer is weak, thin or laterally heterogeneous (e.g. \citefullauthor{olson2018outer} \citeyear{olson2018outer}, \citefullauthor{gastine2020dynamo} \citeyear{gastine2020dynamo}). 

In this paper, we discuss another way to infer and constrain the existence and properties of the stably stratified layer, which involves the wave dynamics in the Earth's core. On short time-scales, perturbations in the convective motions that generate the magnetic field on long time-scales can produce hydromagnetic waves, by which we mean global oscillations of the outer core fluid. The frequencies and other properties of these waves can be used to probe the outer core and, in particular, the possible outermost stable layer. Waves in the Earth's fluid outer core can be (indirectly) observed via either the interaction with the geomagnetic field, resulting in rapid observable changes in the secular variation (SV), which is the first time derivative of the magnetic field \citep[][]{gillet2021dynamical}, or alternatively through angular momentum exchanges between the liquid core and the mantle, leading to observable changes in the length of day (L.O.D.) and polar motions \citep[][]{rekier2021earth}. Four types of wave motions can potentially be identified: (1) torsional Alfvén (TA) waves, (2) Magneto-Coriolis (MC) waves, (3) Magneto-Archimedes-Coriolis (MAC) waves and (4) inertial waves \citep[see][for a review]{triana2021core}. Assuming neutral stratification and quasi-geostrophy \citet[][]{gillet2010fast} identified torsional Alfvén waves and later MC waves \citep[][]{gillet2022satellite} in 1D core flow models reconstructed from geomagnetic data, and linked the TA waves to a six-year oscillation in the L.O.D. variation. \citet[][]{buffett2014geomagnetic} showed that in a 1D model of a stably stratified layer with vanishing radial velocity at the lower boundary, MAC waves can be associated with a 60-year fluctuation of both the magnetic field and the length-of-day \citepfull{buffett2016evidence}.

Here we study these possible waves in a three-dimensional model of the Earth's outer core, incorporating unlike previous studies on the wave dynamics, both an outer stably stratified layer and the neutrally stratified bulk of the core, and allowing for fluid interactions between the two regions. We numerically calculate the wave motions as the solutions of a linear generalised eigenvalue problem, describing the core as a 3D sphere permeated by a magnetic field with a convectively stable layer at the top. We concentrate on waves with interannual to decadal periods to understand how the interannual torsional Alfvén waves are affected by varying stratification strengths, and to investigate the existence of MAC-like waves on longer periods. 

\citet{cox2016observational} previously evaluated whether the amplitude of the induced magnetic field associated with torsional Alfvén waves can be detected in observations of the Earth's secular variation, while \citet[][]{chiduran2021signatures} conducted a similar evaluation for the magnetic signature of MAC waves. In this study, we assess whether the amplitude of these different waves can be detected in time series of Earth's rotation by computing the viscous and electromagnetic torques they exert on the mantle. We examine possible signatures of axisymmetric waves in L.O.D. variations, building on the work of \citet{gillet2010fast} and \citet{buffett2016evidence}. Additionally, we explore the potential signatures of non-axisymmetric waves in polar motion.

\section[]{Computing the fluid motions in a spherical model of \\* the Earth's core} \label{sec:method}
\subsection{Principal equations} \label{ssec:equations}
The interannual and decadal flows that we are interested in operate on much shorter time-scales than the convective core flows that are thought to be responsible for generating the magnetic field \citep[][]{jones2015thermal}. This means that we can consider the velocity $\bb{v}$, magnetic field flux $\bb{B}$, gravitational acceleration $\bb{g}$, density $\rho$ and pressure $p$ of these faster flows to be small perturbations $\{\bb{u}, \bb{b}, \bb{\nabla}\mathit{\Phi}', \rho', p'\}$ to a steady background state $\{\bb{U}_0, \bb{B}_0, \bb{g}_0, \rho_0 + \bar{\rho}, P_0\}$, as is standard practice \citep[see e.g.][for reviews]{finlay2008waves, triana2021core}:
\begin{align}
\bb{v}(\bb{r}, t) &= \bb{u}(\bb{r}, t) + \bb{U}_0(\bb{r});\\
\bb{B}(\bb{r}, t) &= \bb{b}(\bb{r}, t) + \bb{B}_0(\bb{r});\\
\bb{g}(\bb{r}, t) &= \bb{\nabla}\mathit{\Phi}'(\bb{r}, t) + \bb{g}_0(\bb{r});\\
\rho(\bb{r}, t) &= \rho'(\bb{r}, t) + \rho_0(\bb{r}) + \bar{\rho}; \label{eq:density}\\
p(\bb{r}, t) &= p'(\bb{r}, t) + P_0(\bb{r}).
\end{align}
Here $\rho_0(r)$ is a steady background profile and $\bar{\rho}$ is a constant spatial average of the density. The dynamics of the core are governed by the system of equations describing the time evolution of the perturbed quantities. Due to the complexity in solving these equations, idealisations and approximations have to be used. Even then, analytical solutions exist only for very special cases, and in general, we have to rely on numerical approaches.

Here, we start from a very simple idealised model of the Earth's outer core in which the influence of the magnetic field $\bb{B}$ and spatial variations of density $\rho_0, \rho'$ are neglected. We consider a spherical volume $\mathcal{V}$, filled with a homogeneous, incompressible and viscous fluid in hydrostatic equilibrium that has uniform density $\bar{\rho}$ and kinematic viscosity $\nu$. The solid surface of the sphere at radius $R$ represents the solid mantle that is rapidly rotating with the Earth's mean angular velocity $\boldsymbol{\Omega} = \Omega\hat{\bb{z}}$. As we are primarily interested in the outermost region of the core, we do not consider a solid inner core for numerical convenience. In a reference frame attached to the mantle, the set of linear equations describing the flow perturbation $\bb{u}$ to a hydrostatic background flow that is uniformly rotating with the same angular velocity as the mantle (i.e. $\bb{U}_0 = 0$) is given by:
\begin{align}
\bar{\rho}\partial_t \bb{u} + 2\bar{\rho}\bb{\Omega\times u} &= - \bb{\nabla}\left(p' + \bar{\rho}\mathit{\Phi}'\right) + \bar{\rho}\nu\bb{\nabla}^2\bb{u}; \label{eq:momentum_hd} \\
\bb{\nabla\cdot u} &= 0. \label{eq:incompressible_hd}
\end{align}
We impose no-slip conditions on the surface of the sphere:
\begin{equation}
\left.\bb{u}\right|_{r = R} = \bb{0}. \label{eq:noslip}
\end{equation}
This specific boundary condition is also required to implement the thin-layer approximation for the magnetic field; see below.

Time-periodic solutions to eqs \eqref{eq:momentum_hd} and \eqref{eq:incompressible_hd} subject to boundary condition \eqref{eq:noslip} exist and are called \textit{inertial waves} \citep[see e.g.][]{greenspan1968rotatingfluids}. These are global flow oscillations, restored by the Coriolis force $ 
\bb{F}_\text{C} = -2\bar{\rho}(\bb{\Omega \times u})$, characterised by the appearance of internal shear layers as conical surfaces around the rotation axis. In the limit of vanishing viscosity, explicit analytical solutions are available \citep[][]{zhang2001inertial}, and the inertial waves form a complete set of modes \citepfull{ivers2015enumeration} on which any smooth fluid flow can be decomposed.

The effect of a magnetic field is included by adding the Lorentz force $\bb{F}_\text{L} = (\bb{\nabla \times B})\bb{\times B/\mu}$ to the momentum equation \eqref{eq:momentum_hd} and introducing the induction equation \eqref{eq:induction_mhd} that governs the time evolution of the perturbed magnetic field $\bb{b}$:
\begin{align}
\bar{\rho}\partial_t \bb{u} + 2\bar{\rho}\bb{\Omega\times u} &= - \bar{\rho}\bb{\nabla}P + \bar{\rho}\nu\bb{\nabla}^2\bb{u} + \mu^{-1}\left((\bb{\nabla \times B}_0)\bb{\times b} + (\bb{\nabla \times b})\bb{\times} \bb{B}_0 \right); \label{eq:momentum_mhd} \\
\partial_t \bb{b} &= \bb{\nabla \times} (\bb{u \times B}_0) + \eta \bb{\nabla}^2\bb{b}; \label{eq:induction_mhd} \\
\bb{\nabla\cdot b} &= 0 \label{eq:nomono_mhd}.
\end{align}
Here $\eta = 1/\mu\varsigma$ is the magnetic diffusivity with magnetic conductivity $\varsigma$ and magnetic permeability $\mu$ and we have simplified the notation by introducing the reduced pressure $P = p'/\bar{\rho} + \mathit{\Phi}'$. To accommodate for an electrically conducting layer at the bottom of the mantle, as inferred by studies of nutational resonances \citepfull{buffett2002modeling} thus allowing for the presence of an electromagnetic torque acting between the core and the mantle, we adopt the thin layer approximation (TLA), formulated by \citetfull{roberts2010numerical}. In the TLA, the bottom of the mantle is treated as a thin conducting layer with thickness $\Delta$ and diffusivity $\eta_\mathrm{W} = 1/\mu_\mathrm{W}\varsigma_\mathrm{W}$ in which the magnetic field $\bb{b}_\mathrm{W}$ is governed by $\partial_t \bb{b}_\mathrm{W} = \eta_\mathrm{W}\bb{\nabla}^2\bb{b}_\mathrm{W}$. Outside the layer, the mantle is considered electrically insulating, and the magnetic field $\bb{b}_\mathrm{M}$ there satisfies $\bb{\nabla \times b}_\mathrm{M}/\mu_0 = 0$, with vacuüm permeabilty $\mu_0$. The thin layer boundary conditions are then expressed by requiring continuity across both interfaces:
\begin{align}
\left.\bb{b}\right|_{r = R} = \left.\bb{b}_\mathrm{W}\right|_{r = R} = \lim_{\Delta \rightarrow 0, \; \mu_\mathrm{W}\Delta = \text{cst.}}\left.\bb{b}_\mathrm{W}\right|_{r = R+\Delta} = \lim_{\Delta \rightarrow 0, \; \mu_\mathrm{W}\Delta = \text{cst.}}\left.\bb{b}_\mathrm{M}\right|_{r = R+\Delta}, \label{eq:thinwall} 
\end{align}
where $\mathrm{cst.}$ denotes a constant value. An explicit expression of eq. \eqref{eq:thinwall} for the poloidal and toroidal magnetic field can be found in eqs \eqref{eq:thinwall_pol} and \eqref{eq:thinwall_tor}. The thin layer approximation is only valid when $\Delta$ is considerably smaller than the skin depth of the mantle, i.e. when the electric current is nearly uniform within the thin layer. In practice, this means that the following ratios:
\begin{align}
\varsigma_\text{ratio} = \frac{\Delta}{R}\frac{\varsigma_\mathrm{W}}{\varsigma}, && \mu_\text{ratio} = \frac{\Delta}{R}\frac{\mu_\mathrm{W}}{\mu},
\end{align}
should be sufficiently small. $\varsigma_\mathrm{ratio}$ gives the ratio of conductance $\Delta\varsigma_\mathrm{W}$ in the thin layer (i.e. the capacity of the layer to allow for the flow of electric currents) compared to the conductance $R\varsigma$ in the fluid. Exact values for the conductance and permeability ratio of the Earth's lower mantle are unknown, and we follow \citet[][]{buffett2002modeling} in choosing $\Delta = \mathrm{210~m}$ and $\varsigma \approx \varsigma_\mathrm{W}$. We then have $\varsigma_\text{ratio} = \mu_\text{ratio} = 5 \times 10^{-5}$, within the range of conductance ratios $3\times10^{-5} \sim 2 \times 10^{-4}$ estimated by \citet[][]{gillet2010fast} based on acceptable solutions for the torsional Alfvén waves. Furthermore, this choice for the conductance and permeability ratios makes the TLA valid for all wave periods much longer than three days, when the skin depth equals $210~\mathrm{m}$. This more than covers the periods that we are interested in.

\begin{figure}
\figbox*{6.3in}{3.15in}{\includegraphics[]{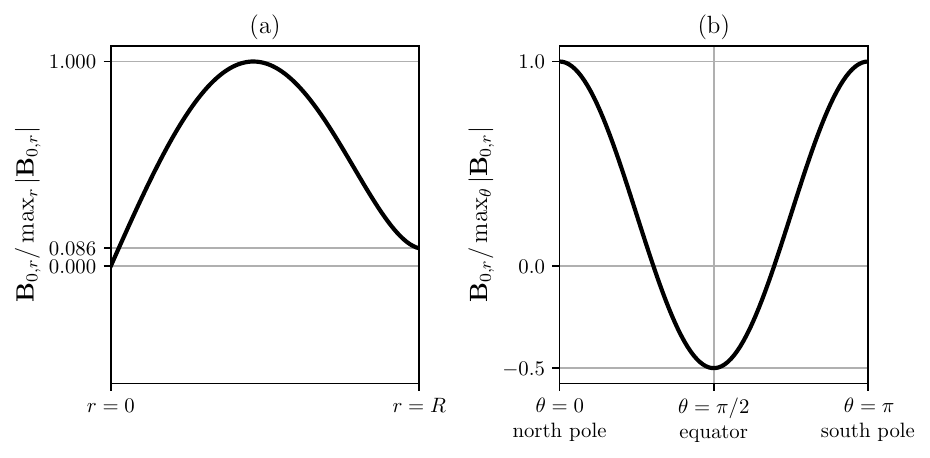}}
\caption{Illustration of the radial (a) and latitudinal (b) dependence of the background magnetic field $\bb{B}_{0,r}$ determined by eq. \eqref{eq:B0}. The left plot shows the field strength $\bb{B}_{0,r}$ as a function of radius $r$ for an arbitrary choice of $\theta,\phi$, divided by its maximum value $\max_r|\bb{B}_{0,r}|$. The right plot shows the field strength $\bb{B}_{0,r}$ as a function of latitude $\theta$ for an arbitrary choice of $r,\phi$, divided by its maximum value. $\max_\theta|\bb{B}_{0,r}|$}
\label{fig:B0_profile}
\end{figure}
Following \citet[][]{luo2022waves} we select a quadrupolar-like background magnetic field that is regular at the origin:
\begin{align}
\bb{B}_0 = \bb{\nabla \times \nabla \times } \left[\frac{B_0}{32} \sqrt{\frac{15}{182\pi}} r^2\left(157 - 296r^2 + 143 r^4\right)\mathrm{Y}_{2 0} \ \bb{r} \right], \label{eq:B0}
\end{align}
where  $B_0^2 = \int_\mathcal{V}|\bb{B}_0|^2d^3\bb{r}$ and $\mathrm{Y}_{2 0}$ is the degree $\ell= 2$ and order $m=0$ Schmidt semi-normalised spherical harmonic with normalisation factor $\sqrt{(\ell-m)!/(\ell+m)!}$. An illustration of this background magnetic field can be found in Fig. \ref{fig:B0_profile}. A quadrupolar field might seem strange considering that the geomagnetic field is found to be mostly dipolar \citep[see e.g.][]{hulot2015present} but the associated Alfvén velocity profile $v_\mathrm{A}(r) = |\bb{B}_0|(r)/\sqrt{\mu\bar{\rho}}$ matches better with the profile that is recovered from wave observations by \citet[][]{gillet2010fast} in the Earth's outer core,  see fig. 2 in \citet[][]{luo2022waves}. Moreover, the strength of the radial component of this background magnetic field at the CMB is a factor 10 weaker than the radial field deep inside the core, see Fig. \ref{fig:B0_profile}(a), as suggested by geodynamo models \citep[e.g.][]{aubert2009modelling}.

The incompressible MHD equations, i.e. \eqref{eq:momentum_mhd}, \eqref{eq:induction_mhd}, \eqref{eq:incompressible_hd}, and \eqref{eq:nomono_mhd}, admit two branches of periodic solutions called \textit{Magneto-Coriolis (MC) waves} (e.g. \citeauthor{malkus1967hydromagnetic} \citeyear{malkus1967hydromagnetic}; \citefullauthor{gerick2021fast} \citeyear{gerick2021fast}) that arise from an interplay between Lorentz and Coriolis forces: (1) slow MC waves when these restoring forces oppose each other and (2) fast MC waves when they work in tandem. In the limit where the Coriolis force dominates, the alignment of the two forces is irrelevant and these waves are essentially inertial waves. In the opposite limit, where the Lorentz force dominates, we find \textit{Alfvén waves}, time-periodic solutions to the MHD equations with $\bb{\Omega} = 0$ that are restored by magnetic tension. In the rapidly rotating Earth's core where the Coriolis force is not negligible on all time-scales Alfvén waves only exist in those geometries within the fluid system where the Coriolis force disappears, i.e. on axisymmetric cylinders that are aligned with the rotation axis $\hat{\bb{z}}$. In that case, the waves appear as axisymmetric oscillations of rigid cylindrical surfaces about that axis and are called \textit{torsional Alfvén (TA) waves} \citep[e.g.][]{braginskiy1970torsional,gillet2010fast}.

Finally, we use the Boussinesq approximation to include variations in density as is common in Earth's outer core studies \citep[see e.g.][]{triana2021core}. We assume that variations in density result from thermal effects, a deviation of the temperature from an adiabatic reference state, chemical effects, a deviation of the light element concentration from a well-mixed fluid (which we will call composition), or a combination of both. We further assume that the combination of thermal and chemical effects can be described by a single variable $C$, the co-density, defined by a single volume coefficient of thermal and chemical expansion $\alpha$ and a single diffusivity $\kappa$. In our Boussinesq system, the equation of state is then given by:
\begin{align}
\rho(\bb{r}, t) &= \bar{\rho} -\alpha\bar{\rho}(C(\bb{r}, t)-\bar{C}); \label{eq:EOS_density}\\
C(\bb{r}, t) &= \bar{C} + C_0(\bb{r}) + \mathit{\Xi}(\bb{r}, t) \label{eq:EOS_anomaly},
\end{align}
where $C$ is the co-density, with a constant spatial average $\bar{C}$, steady background profile $C_0$ and perturbation $\mathit{\Xi}$. In the Boussinesq approach, density perturbations are neglected in the momentum equation \eqref{eq:momentum_mhd}, except when they are multiplied with the gravitational acceleration in the buoyancy force $\bb{F}_\text{B} = \rho\bb{g}$: 
\begin{align}
\bar{\rho}\partial_t \bb{u} + 2\bar{\rho}\bb{\Omega\times u} &= - \bb{\nabla}P + \bar{\rho}\nu\bb{\nabla}^2\bb{u} + \frac{1}{\mu}\left((\bb{\nabla \times B}_0)\bb{\times b} + (\bb{\nabla \times b})\bb{\times} \bb{B}_0 \right) + \frac{\alpha g_0\bar{\rho}}{R}\mathit{\Xi}\bb{r} ; \label{eq:momentum_d} \\
\partial_t \mathit{\Xi} &= \bb{u} \cdot \partial_r C_0 \hat{\bb{r}} + \kappa \bb{\nabla}^2 \mathit{\Xi}. \label{eq:thermal_d} 
\end{align}
Here $\kappa$ is the thermal and chemical diffusivity and both the background gravitational acceleration $\bb{g}_0 = -g_0\bb{r}/R$ and the co-density gradient $\bb{\nabla}C_0 = \partial_r C_0 \hat{\bb{r}}$ depend only on the radial coordinate. To enforce that the density perturbations cause no additional heat or compositional flux into the mantle, we set uniform heat flux conditions at the CMB:
\begin{equation}
\left.\partial_r \mathit{\Xi} \right|_{r = R} = 0. \label{eq:heatflux}
\end{equation}
\begin{figure}
\figbox*{3.15in}{3.15in}{\includegraphics[]{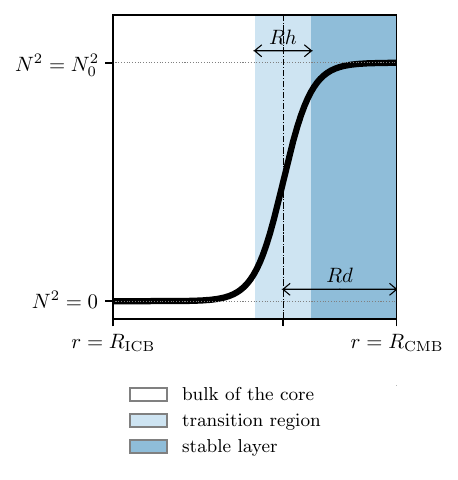}}
\caption{Illustration of the radial profile of the Brunt-Väisälä frequency $N(r)$ determined by the stratification strength $N_0$, smoothness parameter $h$ and layer thickness $d$ in eq.\eqref{eq:BV_profile}.}
\label{fig:BV_profile}
\end{figure}
We let the co-density gradient $\bb{\nabla}C_0 (r)$ be set by the following non-dimensional profile of the Brunt-Väisälä frequency $N^2(r) = \alpha g_0 \bb{\nabla} C_0/\Omega^2$:
\begin{equation}
N^2(r) = \frac{N_0^2}{2\Omega^2}\left(1 + \tanh{\left(\frac{2\left(r - R(1 - d)\right)}{h}\right)}\right). \label{eq:BV_profile}
\end{equation}
With this description of the Brunt-Väisälä frequency \citep[similar to][]{vidal2015quasi}, the convectively unstable $(N^2 \lesssim 0)$ bulk of the core is considered as a region of neutral stratification ($N^2 = 0$) with approximate radius $R(1-d)$ that transitions smoothly, for large values of $h$, or sharply, for small values of $h$, into a region that is stable against convection with $N^2 = N_0^2 > 0$ (Fig. \ref{fig:BV_profile}). The added buoyancy force in eq. \eqref{eq:momentum_d} introduces an additional type of periodic wave motion, typically called a \textit{Magneto-Archimedes-Coriolis (MAC) wave} \citep[e.g.][]{braginsky1993mac, buffett2019equatorially}, that originates from an interplay balance between magnetic (Lorentz), Archimedes (buoyancy) and Coriolis forces. In the same limit where, without a stable layer, fast MC waves become inertial waves, MAC waves are typically called \textit{gravito-inertial waves} (\citefullauthor{dintrans1999gravito} \citeyear{dintrans1999gravito}; \citeauthor{vidal2015quasi} \citeyear{vidal2015quasi}) as they are restored by a combination of the buoyancy and the Coriolis force there. 

In the context of the Earth's outer core, buoyancy-driven waves have already been studied on decadal periods in a fully stratified thin spherical shell with vanishing radial velocity at the lower boundary \citep[][]{buffett2019equatorially, jaupart2017generation, nicolas2023excitation}, or on very short periods, a couple of weeks or less, in a spherical shell with a top stable layer \citep[][]{vidal2015quasi}. Here we study the waves on both decadal and interannual periods, for varying values of the Brunt-Väisälä frequency $N_0$ at the core-mantle boundary while allowing for a flow into and out of the stable layer. We do so by numerically solving a dimensionless version of eqs \eqref{eq:momentum_d}, \eqref{eq:induction_mhd}, \eqref{eq:thermal_d}, \eqref{eq:incompressible_hd}, \eqref{eq:nomono_mhd}, subject to the boundary conditions \eqref{eq:noslip}, \eqref{eq:thinwall}, and \eqref{eq:heatflux}. 

\subsection{Dimensionless equations and parameters} \label{ssec:dimensions}
We choose the core radius $R = \mathrm{3480~km}$ \citep[][]{kennett1995constraints} as the characteristic length-scale, the average magnetic field intensity $B_0 = \mathrm{4~mT}$ \citep[][]{gillet2010fast} as the characteristic magnetic scale, the Alfvén time $\tau_\mathrm{A} = R\sqrt{\mu\bar{\rho}}/B_0 \approx \mathrm{3~yr}$, as the characteristic time-scale  (with vacuum permeability $\mu = 4\pi \times 10^{-7}~\mathrm{H~m^{-1}}$ and average core density $\bar{\rho} = 10^{4}~\mathrm{kg ~m^{-3}}$) and $R\ \partial_r C_0$ as a characteristic scale for the co-density. We then obtain a non-dimensional version of the linearised Boussinesq MHD equations \citep[see e.g.][]{davidson2001magnetohydrodynamics}, when dividing the momentum equation \eqref{eq:momentum_d} by $\bar{\rho}$:
\begin{align}
\partial_t \bb{u} + \frac{2}{\mathrm{Le}}(\hat{\bb{z}}\bb{\times u}) &= -\bb{\nabla} P + \frac{\mathrm{Ek}}{\mathrm{Le}}\bb{\nabla}^2\bb{u} + (\bb{\nabla \times B}_0)\bb{\times b} + (\bb{\nabla \times b})\bb{\times B}_0 + N^2(r)\frac{1}{\mathrm{Le}^2}\mathit{\Xi}\bb{r}; \label{eq:momentum}\\
\partial_t \bb{b} &= \bb{\nabla \times}(\bb{u \times B}_0) + \frac{\mathrm{Em}}{\mathrm{Le}} \bb{\nabla}^2\bb{b}; \label{eq:induction} \\
\partial_t \mathit{\Xi} &= \bb{u \cdot} \hat{\bb{r}} + \frac{\mathrm{Ek}}{\mathrm{Pr}\mathrm{Le}}  \bb{\nabla}^2\mathit{\Xi}; \label{eq:thermal} \\
\bb{\nabla \cdot u} =0; \label{eq:incompressible} \\
\bb{\nabla \cdot b} =0. \label{eq:nomono} 
\end{align}
This introduces the following time-scale ratios:
\begin{align}
\mathrm{Ek} &= \frac{\tau_\Omega}{\tau_\nu} = \frac{\nu}{\Omega R^2}; \label{eq:Ek} \\
\mathrm{Le} &= \frac{\tau_\Omega}{\tau_\mathrm{A}} = \frac{B_0}{\Omega R \sqrt{\bar{\rho}\mu}}; \label{eq:Le} \\
\mathrm{Em} &= \frac{\tau_\Omega}{\tau_\eta} = \frac{\eta}{\Omega R^2}; \label{eq:Em} \\
\mathrm{Pr} &= \frac{\tau_\kappa}{\tau_\nu} = \frac{\nu}{\kappa}; \label{eq:Pr} \\
\widetilde{N} &= N(R) = \frac{N_0}{\Omega}. \label{eq:BV} 
\end{align}
The Ekman number $\mathrm{Ek}$ parametrizes the ratio between the rotation time-scale $\tau_\Omega = \Omega^{-1}$ and the viscous diffusion time $\tau_\nu = R^2/\nu$. The smaller its value, the smaller the influence of viscous forces is compared to the Coriolis forces associated with rotation. For the Earth, this is believed to be very small, $\mathrm{Ek} \approx 10^{-15}$ \citep[][]{olson2015core}, and it is tempting to reduce the computational demand by setting it to zero. In our case, however, we require viscosity as the solutions for smooth diffusionless MAC waves are not guaranteed to exist \citep[see the discussion on stratification effects in][]{triana2021core}. The Lehnert number $\mathrm{Le}$ parametrizes the ratio between the rotation time-scale and the Alfvén time. Expressed differently, it determines the propagation speed of inertial waves compared to the speed of Alfvén waves. Its value in the Earth's outer core is estimated as $10^{-4}$ \citep[][]{gillet2021dynamical}, meaning that inertial waves are thought to travel much faster than Alfvén waves. The magnetic Ekman number $\mathrm{Em}$ is similar to the Ekman number in that it parametrizes the ratio of the rotation time-scale with a diffusion time, here magnetic diffusion, $\tau_\eta = R^2/\eta$. The magnetic diffusion time is expected to be faster than the viscous diffusion time in the Earth's outer core so this number is expected to be larger than the Ekman number, somewhere around $10^{-9}$ \citep[][]{olson2015core}. The Prandtl number $\mathrm{Pr}$ is the ratio of the viscous diffusion time to either the thermal and chemical diffusion time, $\tau_\kappa = R^2/\kappa$. When density variations are dominated by thermal effects it is typically close to or smaller than $1$. When density variations are dominated by compositional effects (in which case it is referred to as the Schmidt number $\mathrm{Sc}$) it is a couple orders of magnitude larger, somewhere between $25$ and $1000$ \citep[][]{bouffard2019chemical}. Finally, the stratification strength $\widetilde{N}$ gives the Brunt-Väisälä frequency evaluated at the top of the outer core compared to the rotation frequency. If it is much smaller than one, the stable layer is said to be weakly stratified, and if it is much larger than one, the layer is said to be strongly stratified. Since both the existence and formation mechanism of the stratified layer are still debated, various $\widetilde{N}$ values have been proposed. Models based on geomagnetic observations \citep[e.g.][]{yan2018sensitivity, buffett2016evidence, olson2018outer} typically favour small to moderate values of the Brunt-Väisälä frequency $0.1 \lessapprox \widetilde{N} \lessapprox 1$ in a thin layer $d \ll 0.06$  while models based on seismic observations \citep[e.g.][]{helffrich2010outer, kaneshima2018array} support larger values of stratification $\widetilde{N} \approx 10$ in thicker layers $d \leq 0.12$. A summary of the typical values of the dimensionless numbers in Earth's outer core can be found in Table \ref{tab:parameters}.

\subsection{Numerical method} \label{ssec:numerics}
We use a spectral method to transform the set of partial differential equations \eqref{eq:momentum}--\eqref{eq:thermal}, solenoidal constraints \eqref{eq:incompressible}, \eqref{eq:nomono}, and boundary conditions \eqref{eq:noslip}, \eqref{eq:thinwall}, \eqref{eq:heatflux} into a generalised numerical eigenvalue problem whose eigensolutions are periodic waves with complex eigenfrequency $\lambda = \sigma + \mathrm{i}\omega$, where $\sigma$ is the non-dimensional wave damping and $\omega$ the non-dimensional frequency. Scalar variables, such as $\mathit{\Xi}$ and $P$, are expanded on a basis of Schmidt semi-normalised spherical harmonics $\mathrm{Y}_\ell^m$ and Chebyshev polynomials of the first kind $\mathrm{T}_k$ as:
\begin{equation}
f(t, r, \theta, \phi) = \mathrm{e}^{\lambda t}\sum_{\ell = 0}^{L}{\sum_{m=-\ell}^\ell{\sum_{k=0}^N{ f_{\ell m}^k \mathrm{T}_k(r)\mathrm{Y}_{\ell m}(\theta, \phi)}}} + \text{c.c.}, \label{eq:expansion}
\end{equation}
assuming a time dependence of the form $e^{\lambda t}$ and adding the complex conjugate $\text{c.c.}$ to keep the variables real. We make use of the divergence-free property ($\bb{\nabla \cdot u} = \bb{\nabla \cdot b} = 0$) of the vector fields $\bb{u}$ and $\bb{b}$ to decompose them in a poloidal and a toroidal potential field \citep[e.g][]{chandrasekhar1961hydrodynamics}:
\begin{align}
\bb{u} = \bb{\nabla \times \nabla \times }(U\bb{r}) + \bb{ \nabla \times }(V\bb{r}) \label{eq:poltor_u}; \\
\bb{b} = \bb{\nabla \times \nabla \times }(F\bb{r}) + \bb{ \nabla \times }(G\bb{r}) \label{eq:poltor_b}, 
\end{align}
where $U$, $V$, $F$, and $G$ are scalar fields that can be further expanded according to eq. \eqref{eq:expansion}. Next we operate with $\hat{\bb{r}} \bb{\cdot\nabla\times}$ and $\hat{\bb{r}} \bb{\cdot\nabla\times\nabla\times}$ on the momentum equation \eqref{eq:momentum} and the induction equation \eqref{eq:induction} to get rid of the pressure $P$. We then obtain for each harmonic degree $\ell$, harmonic order $m$ and Chebyshev degree $k$, five algebraic equations in the five complex coefficients $U_{\ell m}^k$, $V_{\ell m}^k$, $F_{\ell m}^k$, $G_{\ell m}^k$, and $\mathit{\Xi}_{\ell m}^k$, that are subject to the boundary conditions:
\begin{align}
\left.\frac{\mathrm{d}}{\mathrm{d}r}U_{\ell m} \right|_{r = R} = \left. U_{\ell m} \right|_{r = R} = 0; \label{eq:noslip_pol}\\
\left. V_{\ell m} \right|_{r = R} = 0; \label{eq:noslip_tor}\\
\left. \left(1 + \mu_\text{ratio}\ell\right)\frac{\mathrm{d} F_{\ell m}}{\mathrm{d}r} + \frac{\ell}{R}F_{\ell m} + \sigma_\text{ratio} R\left(1 + \frac{\mu_\text{ratio}\ell}{2}\right) \left(\frac{\mathrm{d}^2 F_{\ell m}}{\mathrm{d}r^2} - \frac{\ell(\ell+1)}{R^2}F_{\ell m}\right)\right|_{r = R} = 0; \label{eq:thinwall_pol} \\
\left. G_{\ell m} + \sigma_\text{ratio}R\frac{\mathrm{d} G_{\ell m}}{\mathrm{d}r} \right|_{r = R} = 0; \label{eq:thinwall_tor} \\
\left. \frac{\mathrm{d}\mathit{\Xi}_{\ell m}}{\mathrm{d}r} \right|_{r = R} = 0, \label{eq:noflux} 
\end{align}
where $f_{\ell m} = \sum_{k=0}^N f_{\ell m}^k \mathrm{T}_k(r)$ for each of the five complex coefficients. Eqs \eqref{eq:noslip_pol} and \eqref{eq:noslip_tor} are the no-slip condition \eqref{eq:noslip} in terms of poloidal $U$ and toroidal $V$ scalars \citep[e.g.][]{vidal2015quasi}, while eqs \eqref{eq:thinwall_pol} and \eqref{eq:noslip_tor} give the thin layer condition \eqref{eq:thinwall} in terms of poloidal $F$ and toroidal $G$ scalars \citep[e.g.][]{roberts2010numerical, guervilly2013effect}. Due to the axisymmetry of the background fields $\bb{B}_0$ and $C_0$, we can solve this system of equations for each $m$ independently. The solutions will either be symmetric or anti-symmetric with respect to the equatorial plane, so we can also solve for half of the $\ell$ coefficients at a time (either the odd or the even $\ell$ numbers, depending on the symmetry). Because the quadrupolar background magnetic field $\bb{B}_0$ is equatorially symmetric, $\bb{b}$ has the same equatorial symmetry as $\bb{u}$.

The resulting generalised eigenvalue problem is then given as:
\begin{equation}
A\bb{x} = \lambda B \bb{x} \label{eq:problem}
\end{equation}
where $A$ and $B$ are (sparse) linear operators comprising $N \times 5L/2$ linear equations and $5 L/2$ boundary conditions for $N \times 5L/2$ complex coefficients $\bb{x}$, is then assembled and solved with the numerical tool \textsc{Kore} \citep[][]{triana2022kore}. This publicly available and open-source Python code is based on the very efficient sparse spectral method described in \citet[][]{olver2013fast} and relies on the \textsc{PETSc} \citep[][]{dalcinpazklercosimo2011parallel} and \textsc{SLEPc} library \citepfull{hernandez2005slepc}. Further implementation details are found in \citet[][]{rekier2018inertial} and \citet[][]{triana2019coupling}. 

\begin{table}
\caption{Dimensionless parameter values expected in Earth's outer core and used in our numerical computations.}
\begin{tabular}{lcll}
\hline                                                                                 
Parameter                   & Symbol                    & Proposed outer core values & Numerical value      \\
\hline
Ekman number                & $\mathrm{Ek}$             & $\sim 10^{-15}$                    & $2 \times 10^{-9}$            \\
Lehnert number              & $\mathrm{Le}$             & $\sim 10^{-4}$                     & $10^{-3}$            \\
Magnetic Ekman number       & $\mathrm{Em}$             & $\sim 10^{-9}$                     & $10^{-7}$   \\ 
Prandtl number              & $\mathrm{Pr}$             & $\sim 1 - 10^3$                    & $1$                  \\ 
Stratification strength     & $\widetilde{N}$           & $\sim 0 -10$                       & $\{0.01, \ldots, 10\}$\\
Smoothness parameter        & $h$                       & unknown                            & $0.02$               \\
Stable layer thickness      & $d$                       & $\sim 0 - 0.12$& $0.04$               \\
Conductance ratio           & $\varsigma_\mathrm{ratio}$& $\sim5\times10^{-5}$               & $5\times10^{-5}$     \\       
Permeability ratio          & $\mu_\mathrm{ratio}$      & $\sim5\times10^{-5}$               & $5\times10^{-5}$     \\
\hline
\end{tabular}
\label{tab:parameters}
\end{table}
Solving eq. \eqref{eq:problem} for the exact dimensionless regime expected in Earth's outer core (see Table \ref{tab:parameters}), and in particular for the extremely small Ekman number, is impossible with present-day numerical resources, even with our efficient solver, due to the large spectral resolution required for $N$ and $L$ and more importantly due to round-off errors. We focus instead on a set of magnetic parameters, $\mathrm{Ek} = 2 \times 10^{-9}$, $\mathrm{Em} = 10^{-7} $, $\mathrm{Le} = 10^{-3}$ that is just in reach with our numerical method (using $N = 946$ and $L = 911$), and creates an adequate representation of the hydromagnetic eigenmodes in the Earth's outer core without a stable layer ($\widetilde{N} = 0$), as we will show in Section \ref{ssec:axi}. These parameter values are also comparable to the magnetic parameters used in some numerical geodynamo models \citep[e.g. the Midpath model in][]{aubert2018geomagnetic} and correspond to an Elsasser number $\Lambda = \mathrm{Le}^2 / ~\mathrm{Em} = 10$, similar to the Elsasser number expected in the Earth's outer core. Regarding the unknown thermal and compositional parameters, we fix $\Pr =1$, which could either describe thermal effects with relatively large thermal diffusivity or compositional effects with relatively small diffusivity. For the thickness of the stable layer we choose $d = 0.04$, which matches the value inferred by \citet[][]{buffett2016evidence} to explain observed long-period variations of the geomagnetic field, and is contained in the region with anomalous seismic velocities \citep[][]{kaneshima2018array}. We set $h$ to $0.02$ accordingly, so that the transition region, see Fig. \ref{fig:BV_profile}, falls within our computational domain. Finally to determine how the wave properties are affected by the stratification strength $\widetilde{N}$ we vary the squared Brunt-Väisälä frequency at the CMB from $0.01$, the lower limit inferred by models based on geomagnetic observations \citep[e.g.][]{buffett2017stochastic, yan2018sensitivity}, up to $100$, the upper limit inferred by models based on seismic observations \citep[e.g.][]{helffrich2010outer}, as discussed in Section \ref{ssec:dimensions}.

\subsection{Viscous and electromagnetic torques} \label{ssec:observations}
When investigating the effect of the stably stratified layer, we focus in particular on the torques $\bb{\Gamma}$, that are acting on the mantle. By virtue of our modelling assumptions (Section \ref{ssec:equations}), and in particular the spherical shape of the outer boundary and the purely radial variations in density, the fluid motions in our model of the outer core can only alter the mantle angular momentum, and subsequently the rotation of the mantle, through the combined action of the viscous torque $\bb{\Gamma}_\nu$ and the electromagnetic torque $\bb{\Gamma}_\eta$. In non-dimensional and linear form, these are written as the following integrals over the spherical CMB surface $\mathcal{S}$:
\begin{align}
\bb{\Gamma}_\nu &= \frac{\mathrm{Ek}}{\mathrm{Le}} \oint_\mathcal{S} \left(\bb{r \times} (\bb{\nabla u} + \bb{\nabla u}^\mathsf{T})\right) \bb{\cdot} \ \hat{\bb{r}} \ \mathrm{d}S \label{eq:visctorq}; \\
\bb{\Gamma}_\eta &= \oint_\mathcal{S} \left(\bb{r \times b}\right) \left(\bb{B}_0 \ \bb{\cdot} \ \hat{\bb{r}}\right) \ \mathrm{d}S + \oint_\mathcal{S} \left(\bb{r \times} \bb{B}_0\right)\left( \bb{b \ \cdot} \ \hat{\bb{r}}\right) \ \mathrm{d}S ; \label{eq:emagtorq} 
\end{align}
where $\bb{\nabla u}^\mathsf{T}$ denotes the transpose of the velocity gradient. Following the strategy of Section \ref{ssec:numerics} we can expand the torques on to a sum of Schmidt semi-normalised spherical harmonics $\mathrm{Y}_{\ell m}$ with expansion coefficients $U_{\ell m}(r)$, $V_{\ell m}(r)$, $F_{\ell m}(r)$ and $G_{\ell m}(r)$ that depend on the radius and describe the poloidal and toroidal parts of respectively the velocity and magnetic fields. Then for each azimuthal order $m$ the torque amplitude is given by:
\begin{align}
(m=0) && \mathit{\Gamma}_\nu &=  \frac{\mathrm{Ek}}{\mathrm{Le}} \frac{16\pi}{3} \bigg(V'_{1 0} - V_{1 0}\bigg);  \\ 
&&  \mathit{\Gamma}_\eta &= \frac{32\pi}{5} \ h \ \bigg(\frac{18}{7} G_{30} - G_{10} \bigg) ;  \\
(|m|=1)&& \mathit{\Gamma}_\nu &= \frac{\mathrm{Ek}}{\mathrm{Le}}  \frac{8\pi}{3} \bigg(V'_{1 m} - V_{1 m}\bigg);  \\ 
&& \mathit{\Gamma}_\eta &= \frac{8\pi}{5}  \sgn{m} \ 3 \sqrt{3} \ \mathrm{i} \ \bigg(w' \ F_{2m}-w \ F_{2 m}'\bigg) + w \ \bigg(G_{1m} + \frac{12\sqrt{6}}{7} \ G_{3m}\bigg); \\
(|m|\geq 2) && \mathit{\Gamma}_\nu &= 0; \\
&& \mathit{\Gamma}_\eta &=0.
\end{align}
Here $w(r) = B_0/32~\sqrt{15/182\pi}~ r^2(157r^2 - 296r^4 + 143 r^6)$ designates the generating function of the background magnetic field, see eq. \eqref{eq:B0}, and all radial functions are evaluated at the CMB surface, with simplified notation $f = f(r = R)$ and $f' = \partial_r f(r)|_{r = R}$. The viscous and electromagnetic torques only depend on three harmonic orders m=0, and $m=\pm 1$. $m=0$ describes a torque about the $\hat{\bb{z}}$-axis, which affects the axial component of rotation, i.e. variations in L.O.D.. $m= \pm 1$ describes torques about an axis in the equatorial plane, which affects the equatorial components of rotation, i.e. polar motion (PM), a movement of the symmetry axes of the Earth about the mean rotation axis. Both torques are only non-zero for equatorially symmetric solutions of $\bb{u}$ as the torques in different hemispheres will cancel each other out for equatorially anti-symmetric solutions.

Since we are dealing with an eigenvalue problem, the amplitude of the solution is entirely arbitrary. Therefore, to extract the amplitude of the electromagnetic torque at the CMB, it is necessary to evaluate the amplitudes in comparison to another quantity. In what follows, we normalise the solutions by the calculated r.m.s amplitude of the induced radial magnetic field at the CMB surface $\mathcal{S}$:
\begin{align}
    \langle b_{r.m.s.}\rangle_{\mathcal{S}} = \sqrt{\frac{1}{4\pi R^2}\oint_\mathcal{S} b_r^2 ~\mathrm{d}S}. \label{eq:norm_br}
\end{align}
Another way to normalise the properties of axisymmetric ($m=0$) and axially invariant waves would be by the peak amplitude of the zonal flow:
\begin{align}
    |u_\phi|_{\max} = \max_{z = 0}|u_\phi(r, \theta, \phi)|, \label{eq:norm_uphi}
\end{align}
a quantity that can be (indirectly) observed through inversion of the core flow from geomagnetic data. \citet[][]{gillet2010fast} estimated the maximum amplitude of the azimuthal velocity \eqref{eq:norm_uphi} of interannual torsional Alfvén waves as $|u_\phi|_{\max} \approx 0.4~\mathrm{km}~\mathrm{yr}^{-1}$ and \citet[][]{cox2016observational} estimated the resulting secular variation of the radial magnetic field component as $\partial b_r / \partial t \approx 40-60~\mathrm{nT}$ at the CMB. When either of these values is used to scale other dimensionless quantities of the TA waves, the outcome is nearly identical. This occurs because these waves are both axisymmetric and independent of $z$, as we will show in the next section. However, as our interest lies in both non-axisymmetric ($m=\pm 1$) and longer period waves, that are not necessarily axially invariant, we choose $\langle b_{r.m.s.}\rangle_{\mathcal{S}}$ as our method of normalisation.

\section{Waves in a sphere with neutral stratification} \label{sec:neutral}
\subsection{Axisymmetric motions} \label{ssec:axi}
We start by computing the hydromagnetic waves in our model of the Earth's outer core without a stable layer, providing for the first time an eigenvalue spectrum for the Earth's outer core waves that includes both viscosity and electromagnetic coupling to the mantle. Assuming that there is no stable layer under the CMB and that the Earth's outer core is neutrally stratified throughout, the motions of the core fluid can be described by a non-dimensional version of the incompressible MHD equations \eqref{eq:momentum_mhd}, \eqref{eq:induction_mhd}, \eqref{eq:incompressible_hd} and \eqref{eq:nomono_mhd} or equivalently eqs \eqref{eq:momentum}--\eqref{eq:nomono} with $N(r) = 0$. 
\begin{figure}
\figbox*{3.15in}{2.75in}{\includegraphics[]{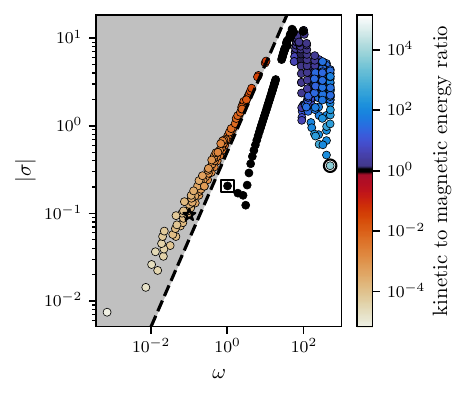}}
\caption{Eigenvalue spectrum of the least-damped equatorially symmetric and axisymmetric ($m=0$) waves in a neutrally stratified core with $\mathrm{Ek} = 2 \times10^{-9}$, $\mathrm{Em} = 10^{-7}$, and $\mathrm{Le} = 10^{-3}$. Frequency $\omega$ on the $x$-axis and damping $\sigma$ on the $y$-axis are non-dimensional with characteristic time-scale $\tau_\mathrm{A}$. Dot colour gives the kinetic to magnetic energy ratio and differentiates between three branches of wave motion: orange for slow Magneto-Coriolis (MC) waves if the magnetic energy is dominant, blue for magnetically modified inertial waves (I) if the kinetic energy dominates and black for torsional Alfvén waves (TA) if the two energies are equal. The dashed line corresponds to $Q = |\omega|/2|\sigma| = 1$, waves above this line in the gray-shaded area have lower quality factors and waves below this line have higher quality factors.} 
\label{fig:kin2mag}
\end{figure}
Fig. \ref{fig:kin2mag} gives the frequency and damping of the least-damped equatorially symmetric solutions to these equations for $m=0$, computed with our numerical method described in Section \ref{ssec:numerics}. We do not determine the full eigenvalue spectrum and as such do not recover all eigenvalues because the size of the matrix makes this a very time-consuming operation and we are not interested in highly dispersive ($|\sigma| \gg |\omega|$) waves that decay before they can be observed. Instead for a series of undamped eigenvalues in our frequency range of interest $\lambda_\mathrm{target} \in \{10^{-3}i, 10^{-2.9}i, \ldots, 10^3i\}$, we use a shift-and-invert method that obtains the $25$ converged solutions with the eigenvalues closest to $\lambda_\mathrm{target}$, depicted as dots in Fig. \ref{fig:kin2mag}. On the black dashed line, the quality factor $Q = |\omega|/2|\sigma|$ is equal to 1.  The quality factor determines the number of periodic oscillations a wave mode undergoes before it gets damped. Above the dashed line, in the grey-shaded area, the wave modes are damped quickly, and are thus less relevant for observations, while below this line the wave modes exhibit several periodic oscillations that could be observed if the mode gets excited. The colour of the dots gives the ratio of kinetic to magnetic energy, which turns out to be a useful quantity to distinguish between different wave types. Indeed, in Fig. \ref{fig:kin2mag} we can identify three qualitatively different branches of wave motions. 

\begin{figure}
\figbox*{6.3in}{3.15in}{\includegraphics[]{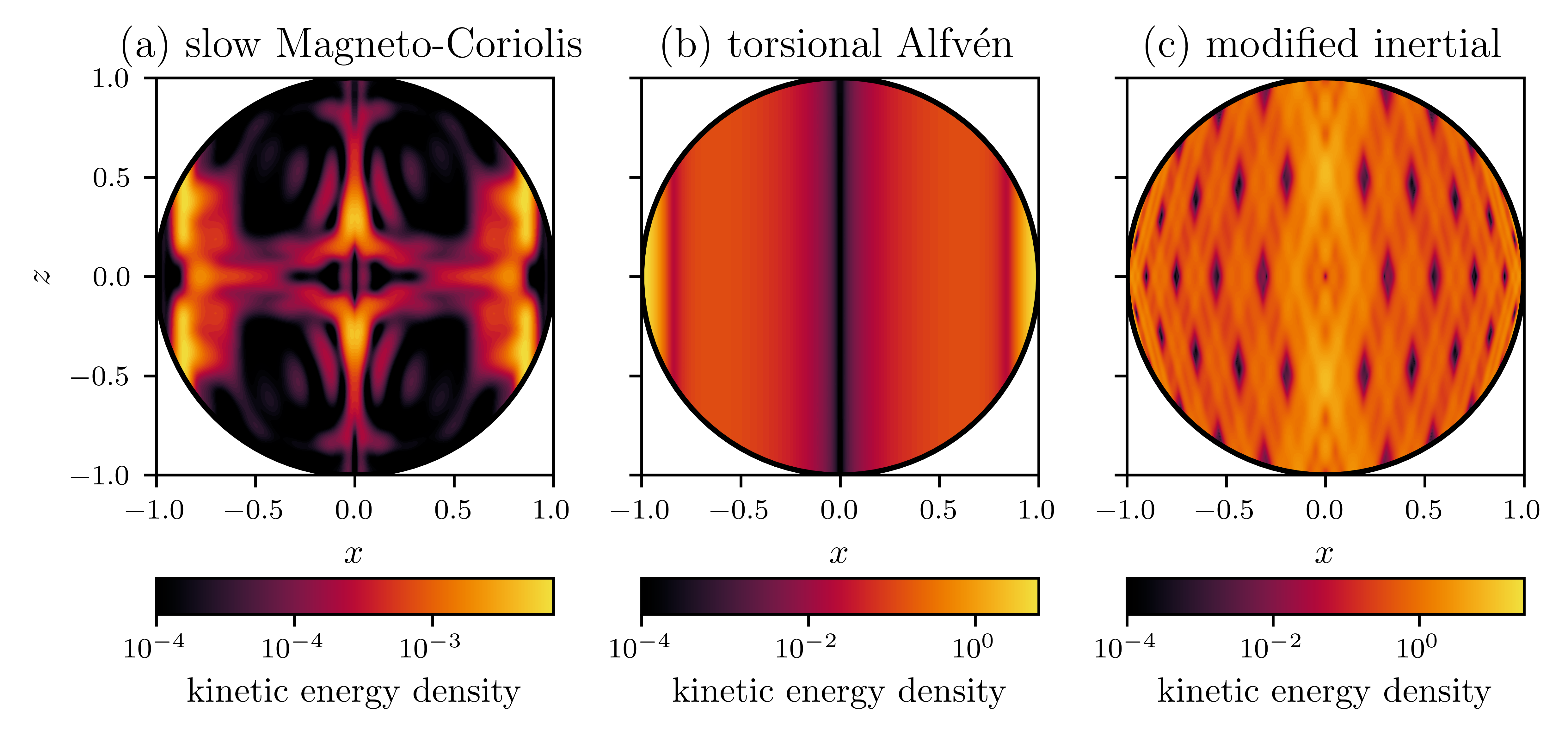}}
\caption{Meridonial cuts of the kinetic energy density in dimensionless units of the waves indicated with a star (a), square (b) and circle (c) in Figure \ref{fig:kin2mag}. Yellow and orange colours indicate regions with higher kinetic energy, while dark and purple colours indicate regions with lower kinetic energy.}
\label{fig:waves}
\end{figure}
High frequency (short period) waves have dominant kinetic energy (Fig. \ref{fig:kin2mag}). An example of the kinetic energy density of one of these waves is shown with a meridional cut at $\phi = 0$ in Fig. \ref{fig:waves}. In this figure, light (orange and yellow) coloured regions correlate with higher kinetic energy, and these zones of peak kinetic energy are organized along internal shear layers, which are characteristic for inertial waves \citep[see e.g.][]{tilgner1999driven}. In the meridonial cut in Figure \ref{fig:waves}(c), these internal shear layers, that form conical surfaces in the full 3D fluid domain, appear as straight lines with a fixed angle to the rotation axis. As the magnetic signature of these waves is small, the waves are too fast to induce significant electric currents \citep[see e.g.][]{braginsky1999dynamics}, we identify them as inertial waves that are slightly modified by the magnetic field, the fast MC waves in Section \ref{ssec:equations}. In the middle of Fig. \ref{fig:kin2mag} where $\mathcal{O}(\omega) \approx 1$ and the wave frequency becomes comparable to the Alfvén frequency, we find a set of wave motions that seem to be invariant along the rotation axis $\hat{\bb{z}}$, see the meridonial cut of the lowest-frequency one in Fig. \ref{fig:waves}(b). These columnar wave motions, not to be confused with convective columns, display an equipartition between kinetic and magnetic energy. These are all distinguishing characteristics of torsional Alfvén waves, and we identify the waves as such. Finally, a set of wave motions with low quality factors and dominant magnetic energy, identifying them as slow Magneto-Coriolis waves, can be found at the left side of Fig. \ref{fig:kin2mag}. These waves are non-columnar for low frequencies, $\omega \ll 1$, see the kinetic energy density plot in Fig. \ref{fig:waves}(a). Such an axial gradient is a distinctive feature of axisymmetric MC waves at long periods \citep[axiMC waves in][]{dumberry2025millennial}. The waves become columnar when their wave period approaches the Alfvén time, $\omega \approx 1$. In a neutrally stratified sphere, permeated by a quadrupolar background magnetic field, their quality factors are below one and they might disperse too quickly to influence the mantle rotation.

The division in TA, MC, and inertial waves in our viscous model is largely comparable to numerical results obtained with inviscid quasi-geostrophic models by \citet[][]{gerick2021fast} and with inviscid models by \citet[][]{luo2022waves}. Only a slight break in the trend of the TA waves, which disappears at lower Ekman numbers, is present. This strengthens our assumption that the set of dimensionless parameter values that we have selected for the flow and magnetic field, see the top rows in Table \ref{tab:parameters}, is suitable for computing the wave motions in an outer core with neutral stratification. 

\begin{figure}
\figbox*{6.3in}{3.15in}{\includegraphics[]{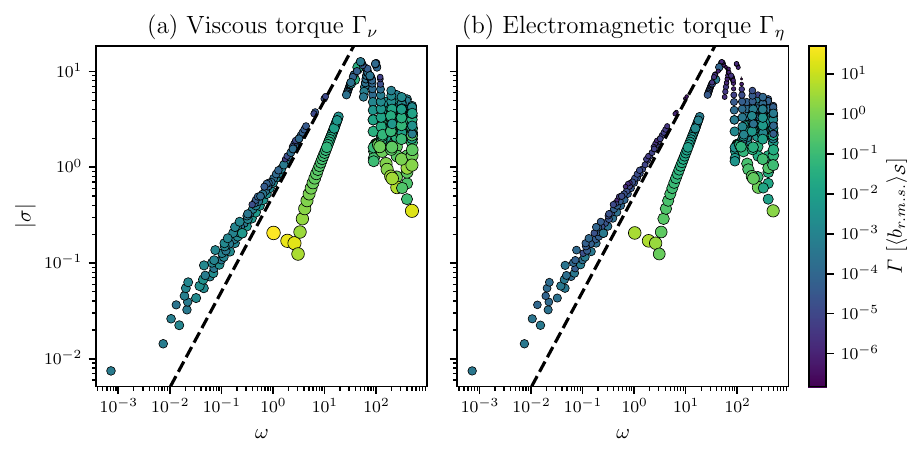}}
\caption{Viscous and electromagnetic torque produced by the least-damped equatorially symmetric axisymmetric waves in a neutrally stratified core with $\mathrm{Ek} = 2 \times10^{-9}$, $\mathrm{Em} = 10^{-7}$, and $\mathrm{Le} = 10^{-3}$. For each wave, depicted with a coloured marker, the torques $\mathit{\Gamma}_\nu$ and $\mathit{\Gamma}_\eta$ have been normalised using the r.m.s value of the wave induced radial magnetic field at the CMB. Light yellow colours and big markers (resp. dark purple colours and small markers) indicate higher normalised torques (resp. lower normalised torques).}
\label{fig:torques}
\end{figure}
Fig. \ref{fig:torques} gives the normalised viscous and electromagnetic torques for the wave spectrum presented in Fig. \ref{fig:kin2mag}. The torques are maximal for the lowest-frequency torsional wave with $\omega = 1.027$ and $\sigma = -0.205$. Using our chosen time-scale, this wave has an approximate period of $P = \tau_\mathrm{A} ~ 2\pi/ \omega = \mathrm{19~yr}$. The viscous torque is about an order of magnitude larger than the magnetic torque. 

The large viscous torque is, however, an artefact of the relatively large Ekman number used in our numerical computations (Table \ref{tab:parameters}). In combination with the steep decline of velocity towards the CMB caused by the no-slip condition and the corresponding large velocity gradient $\bb{\nabla u}$, this dimensionless number overestimates the viscous forces near the boundary. In a previous study \citep[][]{seuren2023effects}, we showed that the viscous torque decreases as $\sqrt{Ek}$, and initial tests show that the same behaviour applies to our numerical results here. Assuming that this is indeed the case, we can make a rough upper estimate of the viscous torque at Earth core conditions:
\begin{align}
\mathit{\Gamma}_\nu = 3\times10^{14}~ \mathrm{N~m}.
\label{eq:ta_vtorq}
\end{align}
This estimate relies on the assumption that torsional waves induce a magnetic signature at the core–mantle boundary of at most $\mathrm{60~nT}$ \citepfull{cox2016observational}. In other words we recover the estimate \eqref{eq:ta_vtorq} by rescaling the flow and magnetic field amplitude so that the induced radial magnetic field at the CMB surface \eqref{eq:norm_br} is equal to $\mathrm{60~nT}$, see also section \ref{ssec:observations}. The L.O.D. variation associated with the torque in \eqref{eq:ta_vtorq} is about $0.45~\mu\mathrm{s}$, see Appendix \ref{app:torque}, much lower than the accuracy of the observed L.O.D. variations \citep[e.g.][]{rekier2021earth}.

The magnetic torque that we compute depends primarily on the r.m.s. radial magnetic field strength at the CMB, captured with the Lehnert number, the dissipation near the core-mantle boundary, captured with the Ekman and magnetic Ekman number, and the unknown conductance ratio $\varsigma_\mathrm{ratio}$ between the mantle and core \citep[see Appendix \ref{app:sensitivity} for the former and][for the latter]{seuren2023effects}. By contrast, the torque is likely only weakly sensitive to the background field morphology \citep[e.g.][]{deleplace2006viscomagnetic, buffett2007magnetic, koot2013role}. In our model, we set $\varsigma_\mathrm{ratio}$ to $5 \times 10^{-5}$ \citep[following][]{buffett2002modeling}, $\mathrm{Ek} = 2\times10^{-9}$ and $\mathrm{Le}=10^{-3}$. With these parameters, the magnetic torque is:
\begin{align}
\mathit{\Gamma}_\eta = 3\times10^{16}~ \mathrm{N~m}
\label{eq:ta_mtorq}
\end{align}
The L.O.D. variation associated with \eqref{eq:ta_mtorq} is approximately $0.05~\mathrm{ms}$. This amplitude is about an order of magnitude smaller than the six-year oscillation in the length-of-day signal, which measures $0.12~\mathrm{ms}$, and can be associated with the lowest-frequency torsional wave \citep[][]{gillet2010fast}, but might be large enough at Earth's core conditions, see Appendix \ref{app:sensitivity}.

A dedicated extrapolation of both torques towards Earth core conditions remains to be done and is beyond the scope of the present study. Rather, our objective is to compare the viscous and electromagnetic torques in a neutrally stratified core to their counterparts in a fluid core with a top stable layer, see Section \ref{sec:stratification}, to understand how stable stratification influences the viscous and electromagnetic coupling between the core and lower mantle. 

Providing estimates for the L.O.D. variation associated with MC waves propagating on longer time-scales is more challenging, as the individual contribution of these waves to the total magnetic field signal is uncertain. We know that on centennial time-scales the total magnetic signal originating from fluid motions in the outer core measures around $\mathrm{100~nT}$ at the Earth's surface, compared to $\mathrm{8~nT}$ at interannual time-scales \citep[][]{lesur2022rapid}. Assuming that the entirety of the magnetic signal is produced by slow Magneto-Coriolis waves, the L.O.D. variation associated with the electromagnetic torque produced by the waves that we find at such periods:
\begin{align}
\Delta\mathrm{LOD} = 1.2~\mu\mathrm{s}, \label{eq:mc_LOD}
\end{align}
which is far below $1.5~\mathrm{ms}$, the amplitude of the 60-year oscillation in L.O.D. \citep[][]{buffett2016evidence}, even taking into account that the actual signal might be slightly larger for lower Ekman numbers. This either indicates that this oscillation in the L.O.D signal is not caused by MC waves in the core or not via the mechanism, electromagnetic coupling at the core-mantle boundary, that we assume here. \citet[][]{more2018convectively, aubert2021interplay} argue for instance that the 60-year signal in L.O.D. may not result from waves in the core at all but rather from forced cylindrical motions in the core. In section \ref{sec:stratification}, we investigate whether or not waves modified by stable stratification in the outermost core could instead be responsible, as proposed by \citet[][]{buffett2014geomagnetic}.

On the shortest time-scales, $P \lesssim \mathrm{1~yr}$ (cf. $\omega \gtrsim 10^{1.5}$ in Figs \ref{fig:kin2mag} and \ref{fig:torques}), the magnetic signal originating from the fluid core will be contaminated with external signals from e.g. the ionosphere and might not be capable of traversing the weakly conducting mantle \citep[][]{gerick2024interannual}. Moreover, on these short time-scales, the L.O.D. signal is mostly driven by changes in atmospheric angular momentum, obscuring possible contributions from fluid motions in the core \citep[see e.g.][]{rekier2021earth}. As such, there is no reliable way to connect the inertial waves that oscillate at these frequencies to the two types of observations that we consider here. We will disregard them in the remainder of the paper and focus instead on the waves with interannual to centennial periods, $10^{-1.5} < \omega < 10^{1.5}$.

\subsection{Non-axisymmetric motions} \label{ssec:non-axi}
\begin{figure}
\figbox*{6.3in}{6.65in}{\includegraphics[]{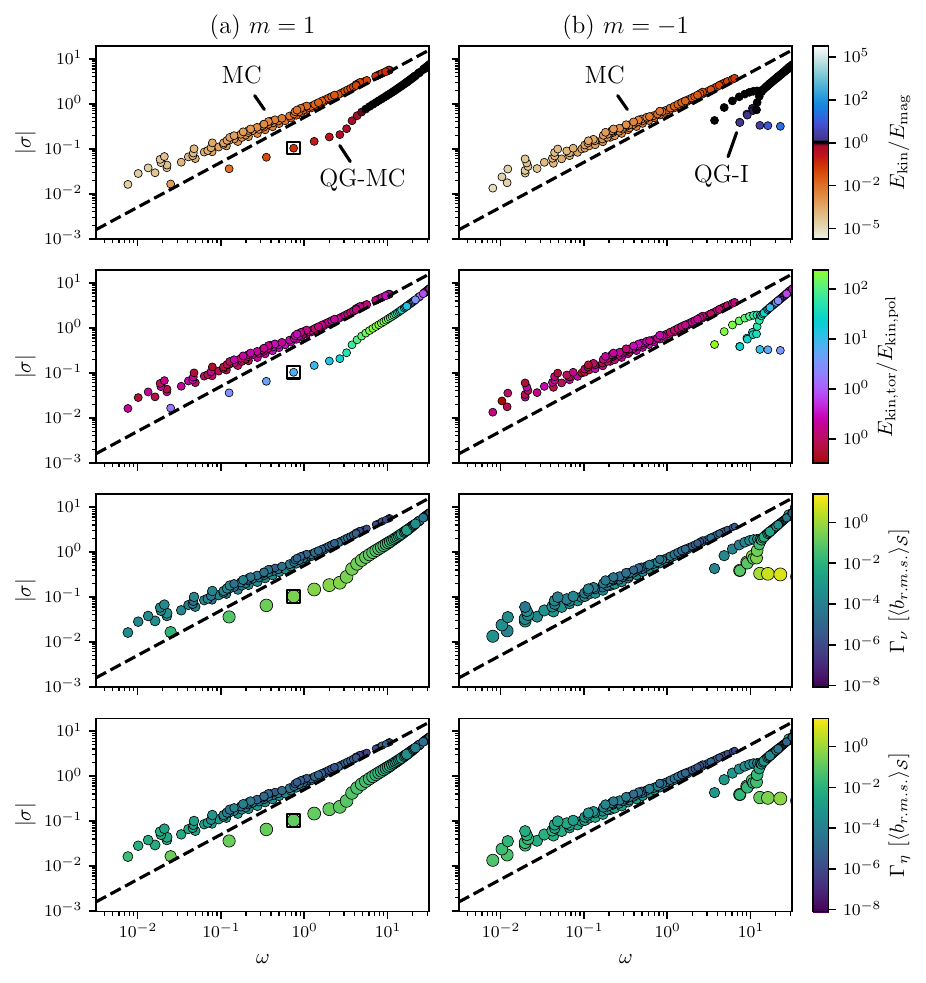}}
\caption{Energy ratios and torques of the least-damped equatorially symmetric eigenmodes within the frequency range $10^{-1.5} \leq \omega \leq 20$ in a neutrally stratified sphere with $\mathrm{Ek} = 2 \times10^{-9}$, $\mathrm{Em} = 10^{-7}$, and $\mathrm{Le} = 10^{-3}$, and with azimuthal orders $m=1$ (a) and $m=-1$ (b). Colour in the top row shows the ratio of kinetic to magnetic energy, where orange corresponds to dominant magnetic energy, blue to dominant kinetic energy, and black to an equipartition, while colour in the second row shows the ratio of toroidal to poloidal kinetic energy, where light blue-green indicates dominant toroidal energy and dark red comparable energies. The third and bottom row show respectively the viscous and electromagnetic torque, where light yellow (resp. dark purple) colours of large (resp. small) dots signify high (resp. low) amplitudes. Branches of slow Magneto-Coriolis waves (MC), columnar Magneto-Coriolis waves (QG-MC) and columnar inertial waves (QG-I) are annotated in the top row. The mode that is closest in frequency to the gravest torsional Alfvén wave (Fig \ref{fig:waves}b) is indicated with a square.}
\label{fig:nonzonal}
\end{figure}
The magnetic field perturbation induced by axisymmetric motions does not explain the full magnetic footprint observed at the Earth's surface (see, for instance, \citet[][]{gillet2022satellite} and the previous Section \ref{ssec:axi}), and so non-axisymmetric waves will also contribute to the observed magnetic signal. Fig. \ref{fig:nonzonal} summarizes our eigenvalue solutions for the non-axisymmetric fluid motions with azimuthal orders $m=1$ (retrograde waves) and $m=-1$ (prograde waves). We can distinguish between two branches of motions. For both harmonic orders, the waves on the left side seem to be slow Magneto-Coriolis waves, similar to the axisymmetric case. Magnetic energy dominates these fluid motions, and meridional cross-sections of their kinetic energy and velocity fields show that their structure is mostly non-columnar, which is supported by the fact that the poloidal kinetic energy is similar to the toroidal kinetic energy. This is in stark contrast with the waves on the right side of the figure that have dominant toroidal energy and, upon visual inspection, are found to have a columnar structure. For high damping, these waves exhibit an equipartition between kinetic and magnetic energy, but when the damping decreases, they become predominantly magnetic for $m=1$ and predominantly kinetic for $m=-1$. Based on these observations, we consider the retrograde waves to be columnar Magneto-Coriolis waves \citep[echoing the quasi-geostrophic (QG-)MC waves in][]{gerick2021fast}, and the prograde waves columnar inertial waves \citep[echoing the QG inertial waves or Rossby waves in][]{gerick2021fast}. 

Compared to the axial torque exerted by the axisymmetric waves, all non-axisymmetric waves exert a substantially lower equatorial magnetic torque on the CMB. For example, for an assumed equal magnetic signal amplitude of $\mathrm{60~nT}$ at the Earth's CMB, the magnetic torque amplitude of the columnar MC wave that is indicated with a square in Fig. \ref{fig:nonzonal} is:
\begin{align}
    \mathit{\Gamma}_\eta = 10^{15}~ \mathrm{N~m}. \label{eq:mc_torq}
\end{align}
This amplitude is lower than those of the lowest-frequency torsional Alfvén wave that is closest in frequency and damping, see eq. \eqref{eq:ta_mtorq}. The actual torque of the MC wave will, of course, depend on its actual magnetic signature, but given that the total magnetic field signal is $\mathrm{8~nT}$, that torque will have a similar order of magnitude. It also remains to be seen if these modes, when extrapolated towards Earth's outer core conditions, can affect the observed polar motion, similar to how axisymmetric motions affect the L.O.D.. Compared to the variations in L.O.D., where, for example, a 6-year periodic signal can clearly be attributed to motions in the fluid core, it is more challenging to extract the core contribution from the polar motion, though some attempts (e.g. \citeauthor{an2023inner}, \citeyear{an2023inner}; \citeauthor{kianishahvandi2024contributions}, \citeyear{kianishahvandi2024contributions}) have been made. This is because the latter signal is dominated by atmospheric, oceanic and hydrospheric forcing, as well as the Chandler Wobble \citep[][]{rekier2021earth}.

\section{Waves in a sphere with an outermost stably stratified layer} \label{sec:stratification}
\begin{figure}
\figbox*{6.3in}{6.65in}{\includegraphics[]{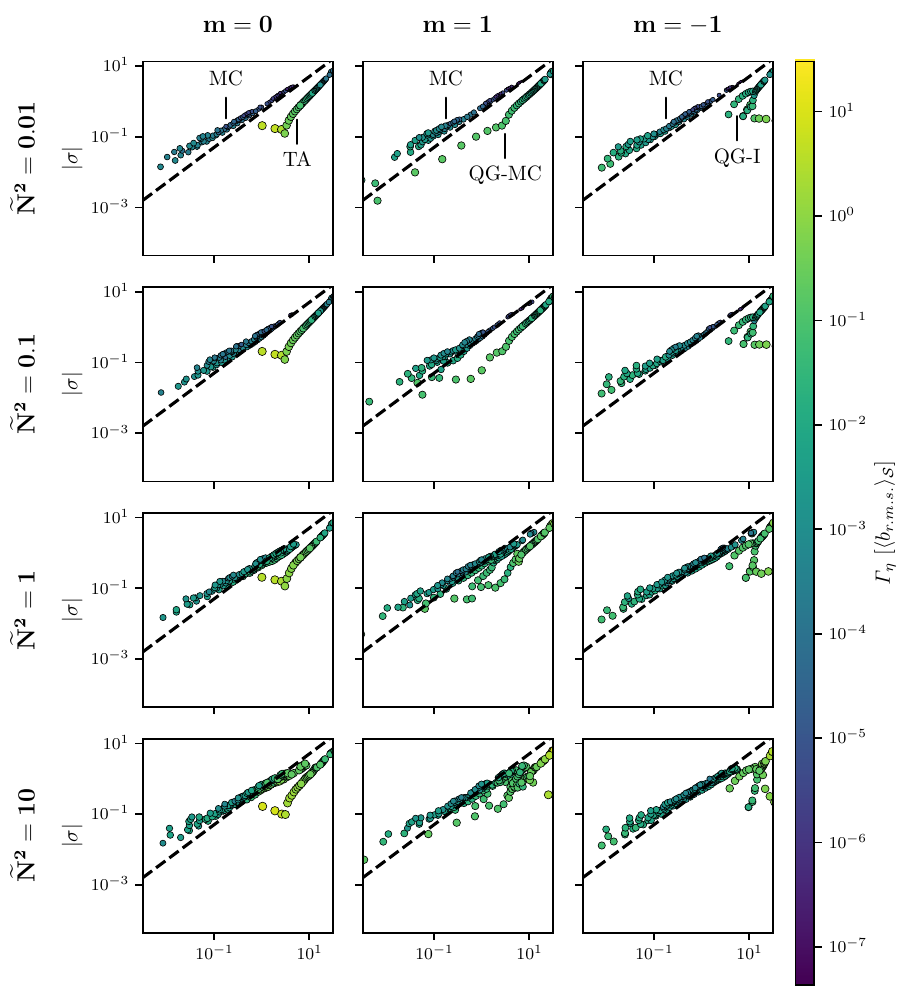}}
\caption{Overview of the electromagnetic torque of the least-damped equatorially symmetric waves in a core with a top stable layer and $\mathrm{Ek} = 2\times 10^{-9}$, $\mathrm{Em} = 10^{-7}$, $\mathrm{Le} = 10^{-3}$. Each dot corresponds to an eigenvalue solution, and its colour and size indicate the magnetic torque strength normalised by the r.m.s. value of the induced radial magnetic field at the CMB. Different columns show different harmonic orders and different rows show different stratification strengths, contrasting very weak ($\widetilde{N}^2 = 0.01$, top row), with weak ($\widetilde{N}^2=0.1$, 2nd row), moderate ($\widetilde{N}^2=1$, 3rd row) and strong ($\widetilde{N}^2=10$, 4th row). The stable layer extends down to about 0.04 of the total spherical radius for each type of stratification. The very weak stratification displayed in the top row has a negligible effect on the eigenvalue spectrum, and the different wave branches are annotated with their wave type in a neutrally stratified sphere.}
\label{fig:spectra}
\end{figure}
The behaviour of the possible fluid motions propagating throughout the Earth's neutrally stratified outer core, as established in Section \ref{sec:neutral}, changes when the outermost region of the core becomes stable against convection. Different types of motions respond differently to the stable stratification which can be seen by comparing the eigenvalue spectra with stratification in Fig. \ref{fig:spectra} to the spectra in a neutrally stratified sphere, Fig. \ref{fig:torques} for $m=0$, Fig. \ref{fig:nonzonal}(a) for $m=1$ and Fig. \ref{fig:nonzonal}(b) for $m=-1$. 

Modes that exhibit an equipartition between kinetic and magnetic energy, i.e. torsional Alfvén waves for $m=0$, high-frequency columnar MC waves for $m=1$ and high-frequency columnar inertial waves for $m=-1$ are the least affected by stable stratification in the outermost region. Noticeable changes to the magnetic torque only occur if the stratification strength exceeds moderate values $\widetilde{N}^2 \gtrsim 1$. The frequency and damping of these modes are mostly unaffected, except for the lowest frequency torsional waves (see further details in Section \ref{ssec:ta}). The very small influence that, in particular, weak stable layers have on the TA waves can probably be attributed to their columnar and toroidal character. We expect the stable stratification to have the strongest effect on the poloidal and radial fluid motions that oppose the radial direction of the co-density gradient in the stable layer \citep[see e.g.][]{seuren2023effects}. The change in frequency and damping of the low-frequency columnar MC waves, with dominant magnetic energy and the low-frequency columnar inertial waves, with dominant kinetic energy, is more complicated, and they are already influenced by weak stably stratified layers, $\widetilde{N}^2 \approx 0.1$. This hints at a more complex interplay between buoyancy, magnetic and rotational forces when the kinetic and magnetic energies are no longer comparable.

The slow Magneto-Coriolis waves with a more poloidal character are affected by weak values of stratification as small as $\widetilde{N}^2 \approx 0.1$. The behaviour of their eigenvalues in the complex spectrum is less smooth than for the torsional Alfvén waves, but on average, they exert a larger magnetic torque on the mantle compared to their neutrally stratified counterparts. The highest frequency waves with increased torques are more interesting from an observational point of view as their quality factor exceeds one. For $m=1$ the effect of the stable layer is so strong that the different wave branches start to mix for the strongest stratified layers $\widetilde{N}^2 \gtrsim 10$ that we consider here.

These general findings show that the relation between the strength of the outermost stable layer and the possible waves that propagate in the Earth's outer core is rather complex, and in the subsections below, we give a more in-depth discussion on this relation by varying the stratification strength in small steps. We narrow our focus to the torsional Alfvén waves (Section \ref{ssec:ta}) and the axisymmetric slow MC waves (Section \ref{ssec:mc}) since these waves, without a stably stratified layer for the former \citep[][]{gillet2010fast}, and without allowing for interaction between the bulk of the core and the stable layer for the latter \citep[][]{buffett2016evidence}, have previously been discussed in the context of rotational variations. 

\subsection{Torsional Alfvén waves} \label{ssec:ta}
\begin{figure}
\figbox*{6.3in}{7.6in}{\includegraphics[]{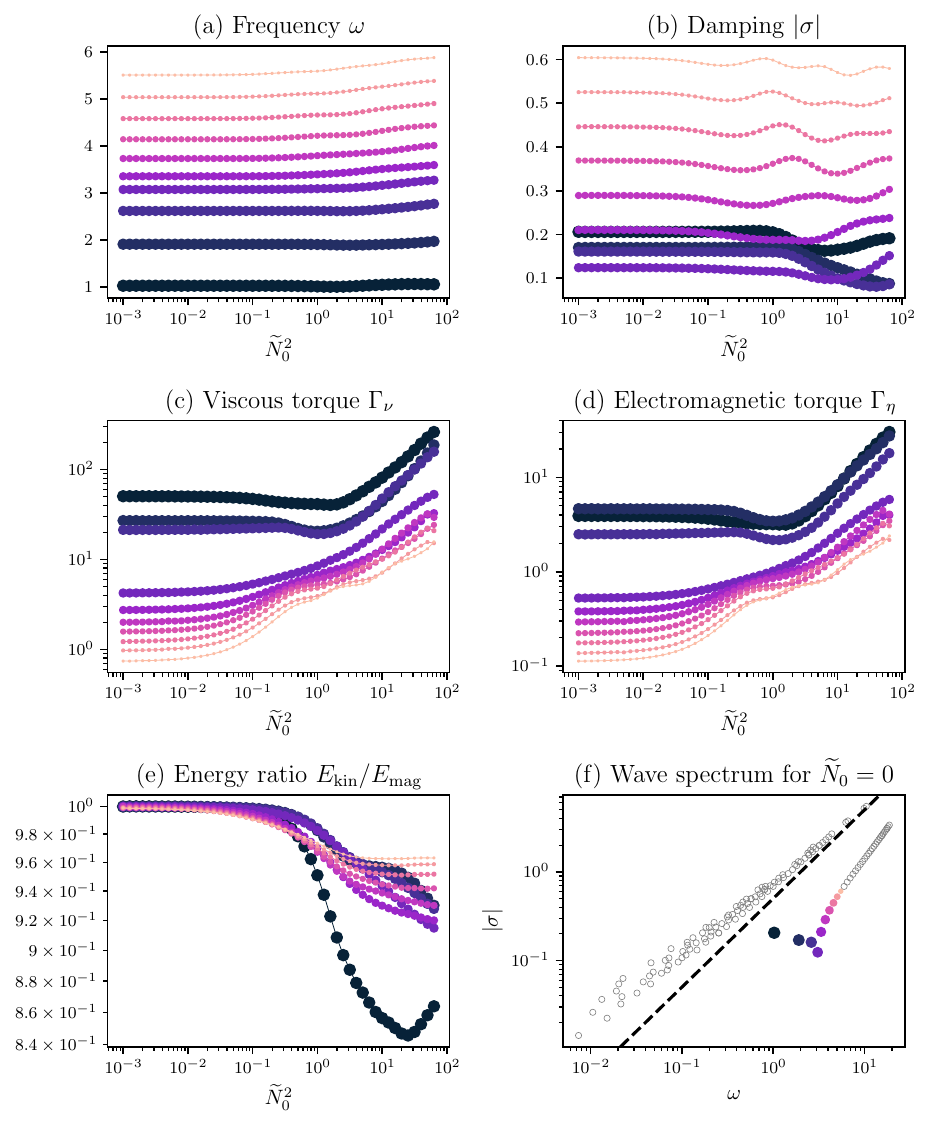}}
\caption{The change of the frequency (a), damping (b), viscous torque (c), electromagnetic torque (d) and kinetic to magnetic energy ratio (e) of the lowest frequency torsional Alvén waves in a sphere with $\mathrm{Ek} = 2 \times 10^{-9}$, $\mathrm{Em} = 10^{-7}$, $\mathrm{Le} = 10^{-3}$, when exponentially increasing the stratification strength of the top stably stratified layer with thickness $d=0.04$ from $\widetilde{N}^2 = 0.01$ towards $\widetilde{N}^{2} = 100$. Marker size and colour indicate the frequency of the 10 lowest frequency torsional Alfvén waves in a neutrally stratified sphere with $\widetilde{N}^2 = 0$ (f).}
\label{fig:ta_varyN}
\end{figure}
The frequency and damping of the torsional Alfvén waves are only marginally affected by increasing the stratification strength $\widetilde{N}^2$ of the top stable layer (see Figs \ref{fig:ta_varyN}a and b). For values below about $N^2 \approx 1$, the frequency is constant and the damping slightly decreases with respect to neutral stratification. For these weak to moderate stratifications, the torques are also only weakly affected, and the waves maintain a near-equipartition of kinetic and magnetic energy. The waves become somewhat more magnetic when increasing the stratification strength (see Fig. \ref{fig:ta_varyN}e). 

\begin{figure}
\figbox*{6.3in}{3.15in}{\includegraphics[]{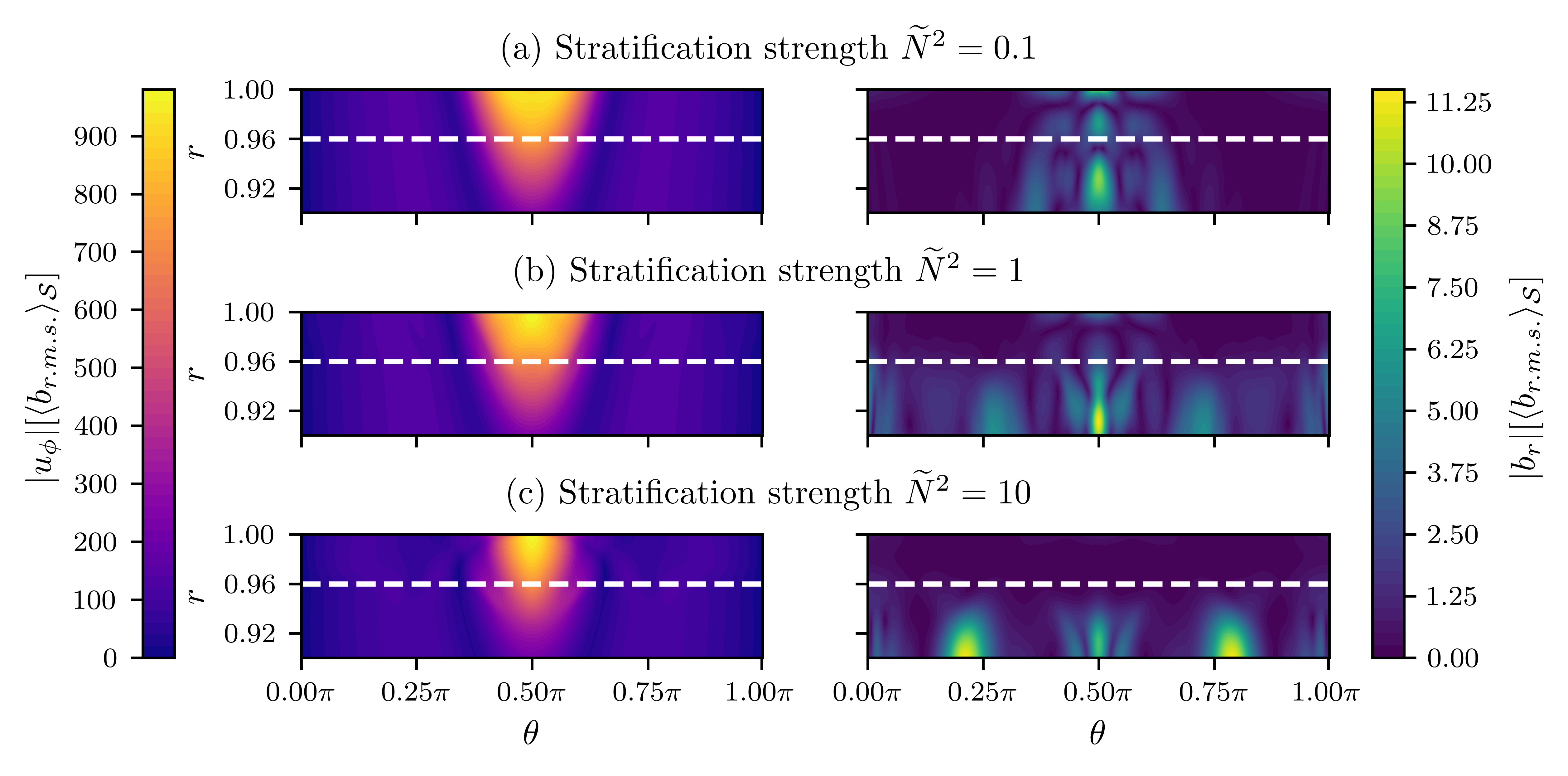}}
\caption{Azimuthal velocity and radial magnetic field near the core-mantle boundary ($0.9 \leq r \leq 1$) of the lowest-frequency torsional wave for different strengths of the stably stratified layer whose lower boundary is indicated by the dashed white lines. Colours express the absolute values of the azimuthal velocity $u_\phi$ (left column) and the radial magnetic field $b_r$ (right column) scaled with the r.m.s. value of the radial induced magnetic field on the CMB surface.}
\label{fig:ta_layer}
\end{figure}

Stronger stratification ($\widetilde{N}^2 > 1$) has a more substantial impact on the frequency, damping and torques (Fig. \ref{fig:ta_varyN}). The velocity and induced magnetic field of the torsional waves are also affected, both in strength and geometry (Fig. \ref{fig:ta_layer}). The azimuthal velocity concentrates more within the stable layer and increases in amplitude when the stratification strength enhances. This directly increases the velocity gradient, augmenting the viscous torque amplitude (Fig. \ref{fig:ta_varyN}c), and also boosts the latitudinal magnetic field (not shown), augmenting the amplitude of the electromagnetic torque (Fig. \ref{fig:ta_varyN}d). Strong stratification in the outer layer, however, significantly diminishes the induced radial magnetic field signal towards the CMB. This means that the increase in the normalised viscous and electromagnetic torque, which are expressed in relation to the r.m.s. value of $b_r$ at the CMB, does not necessarily imply an increase of the actual torque, since the actual induced magnetic field signal at the CMB could be smaller for strong stratification. 

\begin{figure}
\figbox*{3.15in}{3.15in}{\includegraphics[]{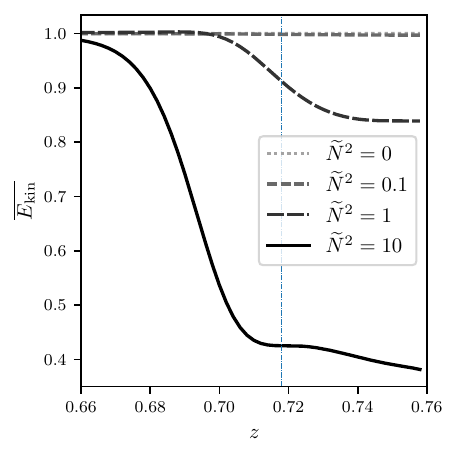}}
\caption{Azimuthally averaged kinetic energy of three torsional waves along the axial line $s = 0.65$ in the upper part of the core $0.66 \leq z \leq 0.76$ where $z \approx 0.76$ is on the CMB. The curves distinguish between the lowest-frequency torsional Alfvén waves for different strengths of the stably stratified layer, with weak stratification ($\widetilde{N}^2 = 0.1$, dotted light grey), moderate stratification ($\widetilde{N}^2 = 1.0$, dashed dark grey), and strong stratification ($\widetilde{N}^2=10.0$, solid black). The curves are scaled by the value of $u_s^2 + u_\phi^2$ in the equatorial plane. The bottom of the stably stratified layer is indicated with the dashdotted blue line.}
\label{fig:penetration}
\end{figure}
A complementary means of evaluating the influence of a stably stratified layer on torsional Alfvén waves is to investigate the extent to which the rigid columnar flows associated with these waves can penetrate the outer stable layer \citep[see, e.g.,][]{takehiro2001penetration, takehiro2002surface}. This is illustrated in Fig. \ref{fig:penetration}, which shows the penetration of such a columnar flow at cylindrical radius $s = 0.65$ into the stable layer. When the outer layer is weakly stratified ($\widetilde{N}^2 = 0.1$), the columnar flow penetrates fully into the stably stratified region, as evidenced by the average kinetic energy remaining essentially unchanged up to the outer boundary. For a moderately stratified layer, $\widetilde{N}^2 = 1$ , the average kinetic energy decreases within the stable region to approximately $0.8$ of its interior value. This reduction is even more extreme in the case of strong stratification ($\widetilde{N}^2 = 10$), where the kinetic energy falls to less than half of its value in the equatorial plane. The effect of both weakly and strongly stratified stable layers does not appear to change when slightly varying the magnetic parameters, see Appendix \ref{app:sensitivity}.

Both Figs. \ref{fig:ta_layer} and \ref{fig:penetration} show that the influence of a weak stable layer on the torsional Alfvén waves is minimal when the stable layer is weakly stratified $\widetilde{N}^2 < 1$. This means that conclusions drawn from the behaviour of the torsional Alfvén waves in a neutrally stratified core are not necessarily in conflict with the existence of a weakly stratified layer. For example, the magnetic field strength in the bulk of the interior, which has been estimated from the frequency of the torsional Alfvén wave \citep[][]{gillet2010fast}, would not change under the influence of a weak stable layer.

\subsection{MAC-like waves} \label{ssec:mc}
\begin{figure}
\figbox*{6.3in}{6.9in}{\includegraphics[]{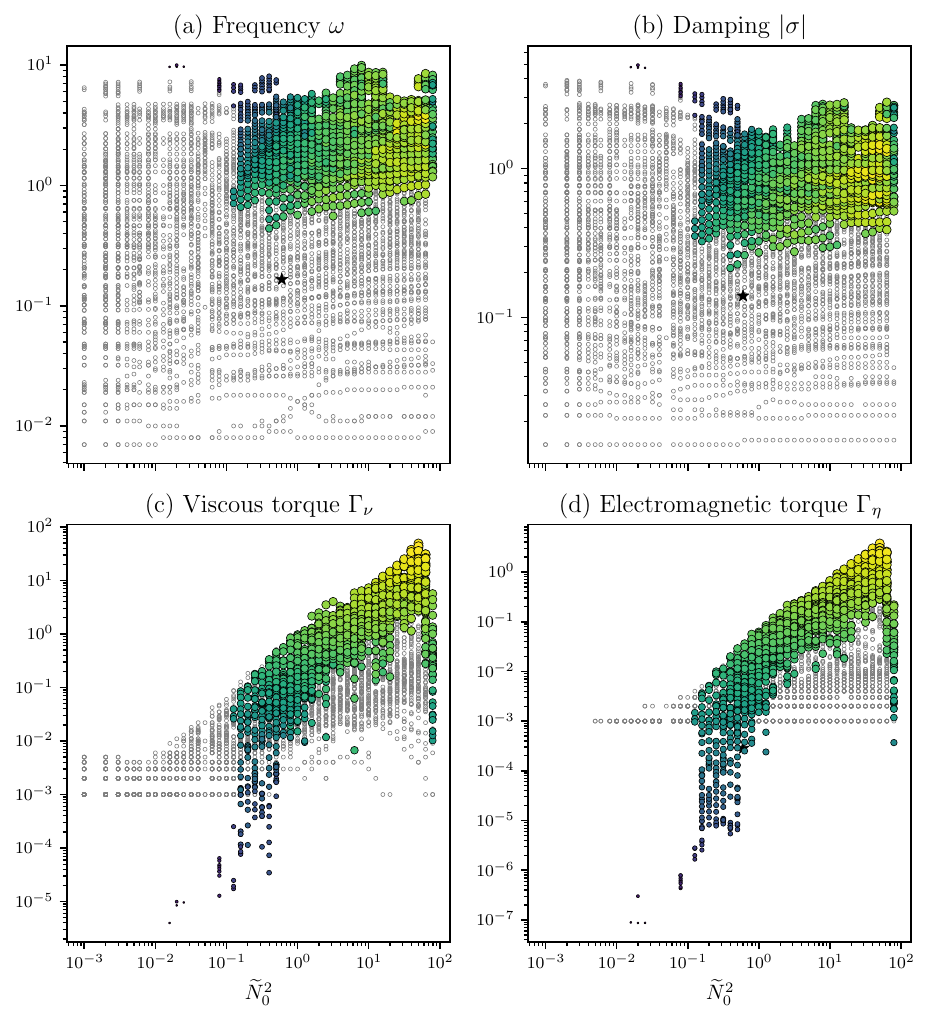}}
\caption{The evolution of the frequency (a), damping (b), viscous torque (c), and electromagnetic torque (d) of the axisymmetric slow MC waves, found in a sphere with $\mathrm{Ek} = 2 \times 10^{-9}$, $\mathrm{Em} = 10^{-7}$, and $\mathrm{Le} = 10^{-3}$, and with a top stably stratified layer of thickness $d = 0.04$ whose squared Brunt-Väisälä frequency ranges from $\widetilde{N}^2 = 0.01$ to $\widetilde{N}^2 = 100$ in a sphere. Fluid motions with quality factor $Q > 1$ are highlighted and marked by coloured disks whose size and colour reflect the magnitude of the magnetic torque that they exert on the CMB, given that the r.m.s. value of the induced magnetic field equals $1~\mathrm{nT}$. All other motions have lower quality factors and are indicated with grey circles. The velocity field of the MAC-like wave designated with a black star can be found in Fig. \ref{fig:compare_Bruce}.}
\label{fig:mc_varyN}
\end{figure}
The complex frequency of the axisymmetric slow MC waves changes in more unpredictable ways than the torsional Alfvén waves, and it is challenging to interpret single mode trajectories when increasing the stratification strength (see the left column of Fig. \ref{fig:spectra}, and the movie in the supplementary information). We can track individual modes through patches of the parameter space after which they seemingly disappear, either because they are no longer converged with our numerical method or because the increasing or decreasing stratification strength no longer supports the modes. It is thus more informative to show, in Fig. \ref{fig:mc_varyN}, a general picture of the evolving eigenvalue spectrum upon incremental changes of the Brunt-Väisälä frequency. This way, we can attribute the influence of the stably stratified layer to a collection of modes.

\begin{figure}
\figbox*{3.15in}{3.15in}{\includegraphics[]{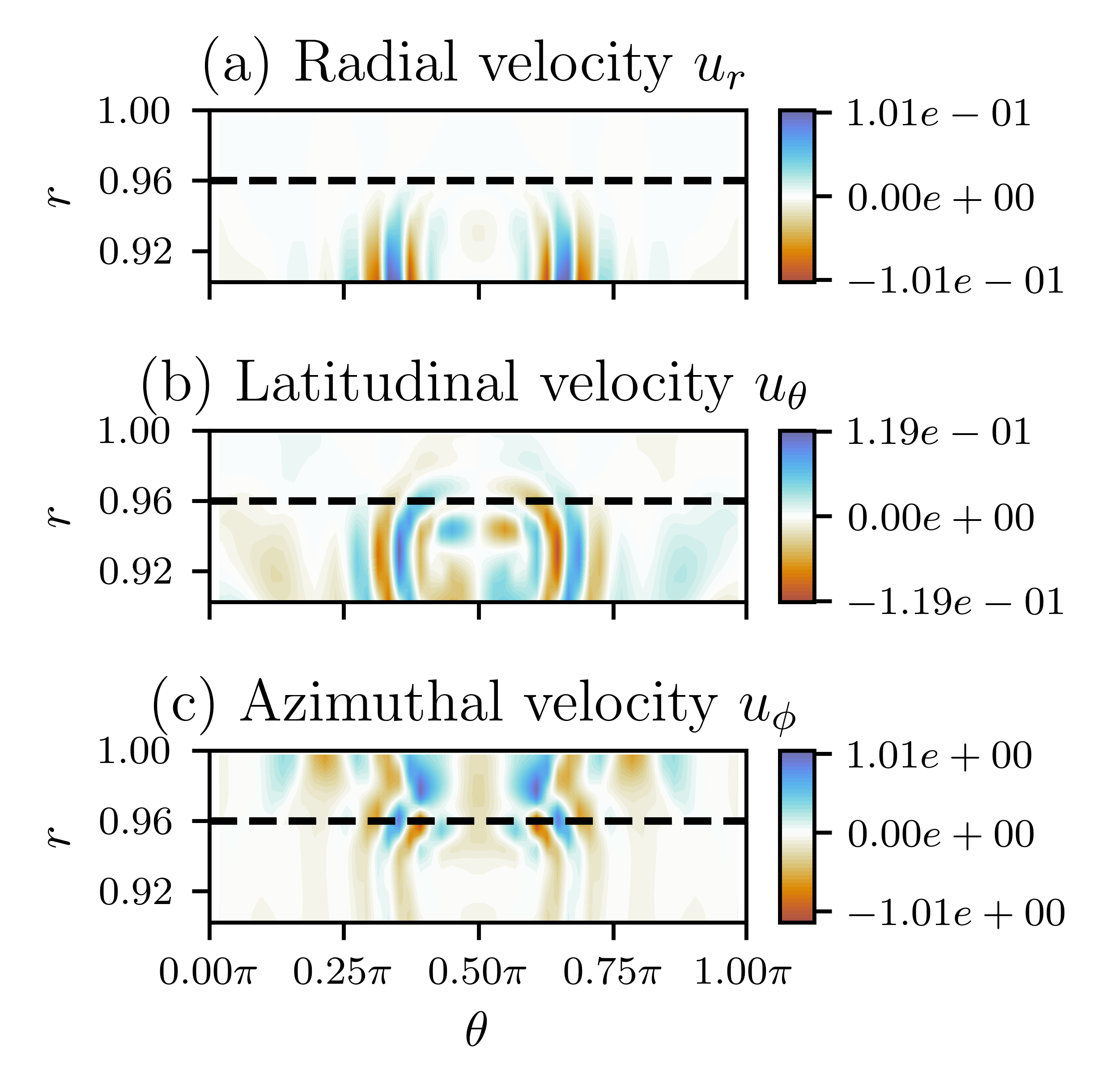}}
\caption{Snapshot of the velocity field of the wave with frequency $\omega=0.166$ and damping $\sigma = -0.140$ (indicated with a black star in Figure \ref{fig:mc_varyN}), within and below a stably stratified layer with thickness $d = 0.04$ and stratification strength $\widetilde{N}=0.6$ at the top of the core ($0.9 \leq r \leq 1$). The lower boundary of the stable layer is indicated by the black dashed line and the top, middle and bottom rows respectively display the radial, latitudinal and azimuthal velocity.}
\label{fig:compare_Bruce}
\end{figure}
For example, up to about $\widetilde{N}^2 \approx 0.1$ the stable layer has a negligible effect on the eigenmodes, as the frequency (Fig. \ref{fig:mc_varyN}a), damping (Fig. \ref{fig:mc_varyN}b), and torques (Fig. \ref{fig:mc_varyN}c and d) all remain about the same. From that point on, the mode behaviour changes dramatically, possibly because the buoyancy force becomes strong enough to affect the dominant force balance between the Lorentz and Coriolis forces. For these, still fairly weak values of the stratification strength, we can find waves that propagate primarily within the stable layer itself, see an example in Fig. \ref{fig:compare_Bruce}. These waves share some similarities with the MAC waves found by e.g. \citet[][]{buffett2016evidence} in a one-dimensional model of a fully stratified thin spherical shell with vanishing radial velocity at the lower boundary. For instance, the ordering of the velocity components, $|v_\phi| > |v_\theta| > |v_r|$,  is identical, with the waves being mainly azimuthal (compare the wave depicted in fig. 2 of \citet[][]{buffett2016evidence} with the MAC-like wave that is nearest in complex frequency and propagates in a stable layer with a comparable stratification strength in Fig. \ref{fig:compare_Bruce}).

There are also notable differences, owing to the increased complexity of our model that (1) allows for the interaction between the stable layer and the neutrally stratified bulk of the core below, (2) has a non-linear profile of the Brunt-Väisälä frequency in the stable layer, and (3) incorporates a quadrupole background magnetic field and electromagnetic coupling between the core and the mantle. In particular, the modes found by \citet[][]{buffett2016evidence} have $Q > 1$, and the azimuthal velocity can be trapped in a region near the equator. \cite{buffett2019equatorially} showed that MAC-waves can become equatorially trapped when the r.m.s. amplitude of the radial magnetic field increases towards the poles.  In our model, where this amplitude is equal to zero at mid-latitudes, maximum at the poles and half the amplitude of the polar value at the equator, see also Fig.\ref{fig:B0_profile}(b), we recover similar trapping behaviour.

It is encouraging that waves confined to the stable layer and trapped near the equator still show up in our more complex model \citep[compared to][]{buffett2014geomagnetic, buffett2016evidence}, and are not, for instance, destroyed by the inflow of radial velocity from the boundary between the stratified layer and the neutral bulk of the core as one might have expected, but could possibly be excited by it. The viscous and electromagnetic torques of these MAC waves can, moreover, be some orders of magnitude larger than the torques of the slow MC waves that are found at nearby frequencies with a similar magnetic field amplitude (see Fig. \ref{fig:mc_varyN}c and d).

\section{Conclusion}
We have used a 3D numerical model to investigate the waves in the Earth's fluid outer core with a convectively stable region beneath the core-mantle boundary, experimenting with different strengths of this stably stratified layer. Compared to previous studies, we included both the interaction between waves in the stable layer and in the neutrally stratified bulk below and the exchange of angular momentum between the core and the mantle. We focused on waves with spherical harmonic orders $m=0,\pm 1$, as only those can influence the mantle's angular momentum through viscous and electromagnetic torques and might consequently be observed in data series of L.O.D. variation and polar motion. We scaled the torque amplitudes and other wave properties according to the r.m.s. value of the radial magnetic field at the CMB surface, another physical quantity that can be linked to observations. Although we compute all kinds of different waves in the core, with and without a stable top layer, our focus is on the influence of the stable layer on the propagation of torsional Alfvén waves and slow MC waves. These are axisymmetric waves that have previously been recovered in geodynamo models \citep[e.g.][]{wicht2010torsional, aubert2021interplay, aubert2022taxonomy} under the assumption of neutral stratification. 

With regards to the torsional Alfvén waves, we showed that these columnar motions, their torques and the magnetic field near the core-mantle boundary are minimally affected by weak to moderate stratification $\widetilde{N}^/\Omega^2 < 1$. This means that proof of their existence is not in direct conflict with the appearance of a stable layer, provided that the stratification in that layer is weak enough. Strongly stratified layers, $\widetilde{N}^/\Omega^2 \gtrsim 10$, affect both the flow structure and the magnetic signature of the Alfvén waves. Most importantly, they substantially reduce the radial magnetic field towards the CMB. This is something that previous studies based on geodynamo models \citep[][]{olson2018outer, gastine2020dynamo} have concluded to be incompatible with the observed geomagnetic field, strengthening the assumption that the Earth's outer stably stratified layer must be weakly stratified if it exists.

The slow MC waves that propagate on interannual to decadal time-scales are already affected by weakly stratified layers, $\widetilde{N}^/\Omega^2 \gtrsim 0.01$. In this frequency range, the weak stably stratified layer allows for the existence of a distinct type of wave that propagates primarily in the stable layer itself. These waves have stronger viscous and electromagnetic torques than the waves in a neutrally stratified core and also share other similarities with so-called MAC waves \citep[e.g.][]{buffett2014geomagnetic, buffett2016evidence}. Like MAC waves, their velocity is dominantly azimuthal and concentrated near the equator. Although these waves experience radial in- and outflow at the lower boundary, which is ignored in the aforementioned studies, these properties are maintained.

Other wave types, such as gravito-inertial or gravity magnetic waves, do not appear in our results and we find no examples for coupling between waves in the stable layer and those in the bulk of the core either. This is likely because the stratification is too weak to support buoyancy-dominated motion on the relatively long time-scales considered. Both phenomena may, however, become relevant on different time-scales, and deserve further study.

There are more aspects of the stable layer than its stratification strength that deserve further investigation, such as its thickness and shape. In particular, it will be very enlightening to examine the influence of the Prandtl number, as this dimensionless number is very different for thermal and chemical effects and can therefore shed light on possible formation (thermal or compositional) mechanisms. In addition, it would be interesting to consider both thermal and compositional effects in a double diffusive approach, where one or both gradients are stable against convection. 

Another future extension of this work would involve investigating the dependence of our numerical results on the choice for a background magnetic field. While we adopted a quadrupolar field for its radial dependence, future studies should assess how our results are modified by a dipolar configuration or by a more realistic geomagnetic field model such as CHAOS-7 \citep[e.g.][]{finlay2020chaos}. The latter would allow for a computation of the secular variation signal associated with the modeled motions. A positive identification of such a signal in geomagnetic data would then constrain the amplitude of these waves, their associated torques, and the resulting changes in LOD and polar motion.

Furthermore, established numerical limitations prevent our numerical method from reaching the exact dimensionless parameter values that are expected in the Earth's outer core. In particular, the high Ekman number tends to overestimate the decay of fluid motions near the outer boundary and consequently the viscous torque. It will be the subject of a further publication to try to establish asymptotic laws for decreasing the Ekman number and determine if our results hold up in the actual dynamical regime of the Earth. Nevertheless, a good agreement between the eigenvalue spectrum computed with our neutrally stratified model at this higher Ekman number and the eigenvalue spectra computed for inviscid models without a stable layer \citep[e.g.][]{luo2022waves} gives us some reassurance that the waves that we compute here are relevant for the fluid motions in the Earth's outer core. 

The results we have presented here are thus a first step in understanding the effect that the suspected stably stratified outer layer has on the waves and electromagnetic torques in the Earth's fluid outer core and show that even when this layer is weak, it can have a powerful effect on the long period motions near the core-mantle boundary and can as such not simply be neglected in studies of the Earth's outer core fluid dynamics. 

\begin{acknowledgments}
The authors would like to thank Mathieu Dumberry and Johannes Wicht for their insightful and helpful comments, which substantially improved the quality of the manuscript. This work has received funding from the European Research Council (ERC) under the European Union’s Horizon 2020 research and innovation program (Synergy grant agreement No. 855677 GRACEFUL) and was financially supported by the Belgian PRODEX program managed by the European Space Agency in collaboration with the Belgian Federal Science Policy Office.
\end{acknowledgments}

\section*{SUPPORTING INFORMATION}
Supplementary data are available at GJI online. 

\textbf{Movie S1.} Evolution of the eigenvalue spectrum and the electromagnetic torque of the least-damped equatorially symmetric and axisymmetric waves in a core with a top stable layer with thickness $d=0.04$ and dimensionless parameter values $\mathrm{Ek} = 10^{-8}, \mathrm{Em} = 5 \times 10^{-7}$, and $\mathrm{Le} = 10^{-3}$. Each dot corresponds to an eigenvalue solution and its colour and size indicate the magnetic torque strength at the core-mantle boundary giving an r.m.s value of the induced magnetic field equal to $1~\mathrm{nT}$. The stratification strength $\widetilde{N}$ increases in exponential steps and is displayed in the top left corner.

\section*{DATA AVAILABILITY}
The code implementing the model described in this study is based on the open source and freely available code \textsc{Kore} \citep[][]{triana2022kore}. All the scripts and associated data necessary to reproduce the figures are archived along with the code at \url{https://doi.org/10.5281/zenodo.14527472}.

\bibliography{bibliography}{}
\bibliographystyle{gji}

\appendix

\section{Rough estimates of the contribution of internal waves to the variations of length-of-day}\label{app:torque}
The axial torque acting on the mantle from flows in the core is related to the change in the rotation rate of the mantle as:
\begin{equation}
    |\mathit{\Gamma}_{z}| = C_\mathrm{m + cr}|\dot{\Omega}_z|
\end{equation}
where $C_\mathrm{m+cr}$ denotes the moment of inertia of the mantle and crust and $|\dot{\Omega}_z|$ the angular acceleration of the mantle around the $z$-axis. The former is estimated as $7.123 \times 10^{37}~ \mathrm{kg~m^2}$ \citep[cf.][]{gross2015earth} and the latter is determined by the period ${\Delta t}_\mathrm{LOD}$ and the average amplitude $\Delta \mathrm{LOD}$ of the observed length-of-day signal:
\begin{equation}
    |\dot{\Omega}_z| = \left(\frac{2\pi}{\mathrm{LOD}_\mathrm{mean} - \Delta \mathrm{LOD}} - \frac{2\pi}{\mathrm{LOD}_\mathrm{mean}}\right)\frac{2\pi}{{\Delta t}_\mathrm{LOD}}
\end{equation}
where $\mathrm{LOD}_\mathrm{mean} = \mathrm{1~d} = \mathrm{86400~s}$ is the mean rotation period of the Earth. For the observed 6-year oscillation signal with ${\Delta t}_\mathrm{LOD} = \mathrm{6~yr}$ and $\Delta\mathrm{LOD} = \mathrm{0.12~ms}$ (\citefullauthor{abarcadelrio2000interannual} \citeyear{abarcadelrio2000interannual}; \citefullauthor{ding2021new} \citeyear{ding2021new}) we find $|\mathit{\Gamma}_{z}| = 2 \times 10^{17} ~ \mathrm{N~m}$ and for the observed 60-year oscillation with ${\Delta t}_\mathrm{LOD} = \mathrm{60~yr}$ and $\Delta \mathrm{LOD} = \mathrm{1.5~ms}$ \citep[][]{buffett2016evidence} we find $|\Gamma_{z}| = 3 \times 10^{17} ~ \mathrm{N~m}$.

Alternatively, the variation in length of day, $\Delta \mathrm{LOD}$, associated with a wave of dimensionless frequency $\omega$, which exerts a torque with amplitude $\mathit{\Gamma}$ on the core-mantle boundary, can be roughly estimated as:
\begin{align}
\Delta\mathrm{LOD}  = \mathrm{LOD}_\mathrm{mean} - 2\pi \left(\frac{\mathit{\Gamma}}{C_\mathrm{m+cr}}\frac{\tau_A}{\omega} + \frac{2\pi}{\mathrm{LOD}_\mathrm{mean}} \right)^{-1},
\end{align}
where $\tau_\mathrm{A}$ is the Alfvén time, the characteristic time-scale that relates the dimensional period of the wave $\Delta t$ to its dimensionless frequency $\omega = 2 \pi \tau_\mathrm{A} / \Delta t $.

\section{Sensitivity of the numerical results to magnetic parameters} \label{app:sensitivity}
While a dedicated extrapolation of our results to the dynamical regime expected in the Earth's outer core is beyond the scope of the present study, this Appendix offers some preliminary insights into how varying magnetic parameters may influence our findings.

In particular, we investigate how the electromagnetic torque, $\mathit{\Gamma}_\eta$, of the lowest-frequency torsional Alfvén wave and its penetration into the stably stratified layer are affected by the Lehnert number $\mathrm{Le}$, and the Ekman $\mathrm{Ek}$ and magnetic Ekman number $\mathrm{Em}$. To quantify the degree of penetration, we define \citep[following][]{vidal2015quasi} a transmission coefficient $\mathcal{C}$, which compares the average kinetic energy just below the viscous boundary layer with the average kinetic energy in the equatorial plane along an arbitrary axial line ($s=0.65$ in Fig. \ref{fig:penetration}) that passes through both the stable layer and the neutrally stratified bulk.
\begin{align}
\mathcal{C}(s) = \frac{\overline{E_\mathrm{kin}(s, z_\mathrm{CMB}(s)-R_\mathrm{CMB}\sqrt{\mathrm{Ek}})}}{\overline{E_\mathrm{kin}(s, 0)}}.
\end{align}
Here, $\overline{\cdot}$ indicates an azimuthal average; $s$ and $z$ are cylindrical coordinates; $z_\mathrm{CMB}(s)$ refers to the height of the core–mantle boundary at a given cylindrical radius $s$ (e.g. $z_\mathrm{CMB}(s) = 0.76$ in Fig. \ref{fig:penetration}); and $R_\mathrm{CMB}\sqrt{\mathrm{Ek}}$ estimates the thickness of the viscous boundary layer. 

\begin{figure}
\figbox*{6.3in}{3.5in}{\includegraphics[]{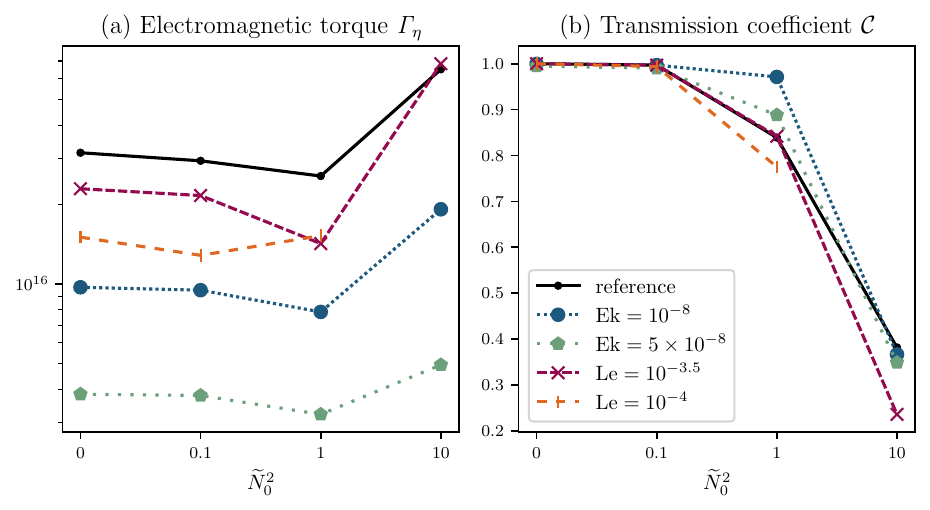}}
\caption{Electromagnetic torque $\mathit{\Gamma}_\eta$ and transmission coefficient $\mathcal{C}$ for different stratification strengths of a top stably stratified layer with thickness $d=0.04$. The solid black line corresponds to a reference solution used in Sections \ref{sec:neutral} and \ref{sec:stratification} with $\mathrm{Ek} = 2\times 10^{-9}$, $\mathrm{Em} = 10^{-7}$ and $\mathrm{Le} = 10^{-3}$. }
\label{fig:sensitivity}
\end{figure}
\subsection{Sensitivity of the magnetic torque}
We consider three values for the Lehnert number, $\mathrm{Le} = 10^{-3}, 10^{-3.5}$ and $10^{-4}$, while keeping the (magnetic) Ekman number constant. This corresponds physically to a situation in which the strength of the background magnetic field varies, while the viscous and magnetic diffusivity remain unchanged. With the possible exception of the strongest stratification case, $\widetilde{N}^2_0$, the magnetic torque decreases slightly when the Lehnert number decreases towards its expected value in the Earth’s core (Fig.\ref{fig:sensitivity}(a)).

Alternatively, we consider three values for the Ekman number, $\mathrm{Ek} = 5\times10^{-8}, 10^{-8}$ and $10^{-9}$, and adjust the magnetic Ekman number accordingly as $\mathrm{Em} = \mathrm{Ek} / 0.02$, while keeping the Lehnert number constant. As the Ekman numbers decrease, the magnetic torque increases. The reduced dissipation in the boundary layer therefore seems to enable a higher magnetic torque, but it remains to be seen whether this trend persists as the Ekman number approaches its expected value in the Earth’s core, $\mathrm{Ek} = 10^{-15}$.

\subsection{Sensitivity of the transmission coefficient}
For all magnetic parameters considered in Fig.\ref{fig:sensitivity}(b), the lowest-frequency torsional Alfvén wave is minimally affected by a weak stable layer with $\widetilde{N}_0^2 = 0.1$, into which it penetrates fully towards the outer boundary. In contrast, across all magnetic parameters, the same wave is significantly affected by strongly stratified layers with $\widetilde{N}_0^2 = 10$, losing more than half its energy in the stable layer. The transition between regimes, where torsional Alfvén waves are either influenced by the stable layer or largely unaffected, thus appears to lie between weak ($\widetilde{N}_0^2 = 0.1$) and strong ($\widetilde{N}_0^2 = 10$) stratification.

\bsp 

\label{lastpage}

\end{document}